\documentclass[aps,prd,twocolumn,groupedaddress,nofootinbib]{revtex4}
\pdfoutput=1
\usepackage{latexsym,amsmath,amssymb,graphicx,booktabs}
\usepackage[english]{babel}

\usepackage{dcolumn}

\usepackage{bm}

\renewcommand\({\left(}
\renewcommand\){\right)}
\renewcommand\[{\left[}
\renewcommand\]{\right]}

\newcommand\n{{\mbox {\boldmath $\nabla$}}}
\newcommand{\ra}{\rightarrow}

\def\lsim{\raise 0.4ex\hbox{$<$}\kern -0.8em\lower 0.62
ex\hbox{$\sim$}}

\def\gsim{\raise 0.4ex\hbox{$>$}\kern -0.7em\lower 0.62
ex\hbox{$\sim$}}

\def\lbar{{\hbox{$\lambda$}\kern -0.7em\raise 0.6ex
\hbox{$-$}}}

\newcommand\eq[1]{eq.~(\ref{#1})}
\newcommand\eqs[2]{eqs.~(\ref{#1}) and (\ref{#2})}
\newcommand\Eq[1]{Equation~(\ref{#1})}

\newcommand\eqst[2]{eqs.~(\ref{#1})--(\ref{#2})}
\newcommand\pa{\partial}
\newcommand\p{\partial}

\newcommand\ee{\end{equation}}
\newcommand\be{\begin{equation}}
\def\bea{\begin{array}}
\def\eea{\end{array}}\def\ea{\end{array}}
\newcommand\ees{\end{eqnarray}}
\newcommand\bees{\begin{eqnarray}}
\def\nn{\nonumber}

\def\d{\delta}

\def\eps{\epsilon}

\def\dslash{\hspace{-1mm}\not{\hbox{\kern-2pt $\partial$}}}
\def\Dslash{\not{\hbox{\kern-4pt $D$}}}
\def\pslash{\not{\hbox{\kern-2.1pt $p$}}}
\def\kslash{\not{\hbox{\kern-2.3pt $k$}}}
\def\qslash{\not{\hbox{\kern-2.3pt $q$}}}

\newcommand{\vac}{|0\rangle}
\newcommand{\cav}{\langle 0|}

\newcommand{\vx}{{\bf x}}

\def\p1{{\bf p}_1}
\def\p2{{\bf p}_2}
\def\k1{{\bf k}_1}
\def\k2{{\bf k}_2}

\newcommand{\emn}{\eta_{\mu\nu}}

\newcommand{\gmn}{g_{\mu\nu}}

\newcommand{\gMN}{g^{\mu\nu}}

\newcommand{\gBmn}{\bar{g}_{\mu\nu}}

\newcommand{\hmn}{h_{\mu\nu}}

\newcommand{\pam}{\pa_{\mu}}

\newcommand{\pan}{\pa_{\nu}}

\newcommand{\Rmn}{R_{\mu\nu}}
\newcommand{\RMN}{R^{\mu\nu}}
\newcommand{\Gmn}{G_{\mu\nu}}

\newcommand{\Tmn}{T_{\mu\nu}}

\newcommand{\TMN}{T^{\mu\nu}}

\newcommand{\dddM}{\kern 0.2em \raise 1.9ex\hbox{$...$}\kern -1.0em \hbox{$M$}}
\newcommand{\dddQ}{\kern 0.2em \raise 1.9ex\hbox{$...$}\kern -1.0em \hbox{$Q$}}
\newcommand{\dddI}{\kern 0.2em \raise 1.9ex\hbox{$...$}\kern -1.0em\hbox{$I$}}
\newcommand{\dddJ}{\kern 0.2em \raise 1.9ex\hbox{$...$}\kern-1.0em
\hbox{$J$}}
\newcommand{\dddcalJ}{\kern 0.2em \raise 1.9ex\hbox{$...$}\kern-1.0em
\hbox{${\cal J}$}}

\newcommand{\dddO}{\kern 0.2em \raise 1.9ex\hbox{$...$}\kern -1.0em
\hbox{${\cal O}$}}
\def\dddz{\raise 1.5ex\hbox{$...$}\kern -0.8em \hbox{$z$}}
\def\dddd{\raise 1.8ex\hbox{$...$}\kern -0.8em \hbox{$d$}}
\def\dddbd{\raise 1.8ex\hbox{$...$}\kern -0.8em \hbox{${\bf d}$}}
\def\ddbd{\raise 1.8ex\hbox{$..$}\kern -0.8em \hbox{${\bf d}$}}
\def\dddx{\raise 1.6ex\hbox{$...$}\kern -0.8em \hbox{$x$}}

\newcommand{\Sch}{Schwarzschild }

\newcommand{\mpl}{M_{\rm Pl}}

\newcommand{\lpl}{l_{\rm Pl}}

\newcommand{\lc}{\Lambda_c}

\newcommand{\px}{\partial_x}

\begin{document}

\title{Zero-point quantum fluctuations  in cosmology}

\author{Lukas Hollenstein}
\author{Maud Jaccard}
\author{Michele Maggiore}
\author{Ermis Mitsou}
\affiliation{D\'epartement de Physique Th\'eorique and Center for Astroparticle Physics, 
Universit\'e de Gen\`eve, 24 quai Ansermet, CH--1211 Gen\`eve 4, Switzerland}

\email{lukas.hollenstein@unige.ch, maud.jaccard@unige.ch, michele.maggiore@unige.ch, ermis.mitsou@unige.ch}

\begin{abstract}
We re-examine the classic problem of the renormalization of zero-point quantum fluctuations in a Friedmann-Robertson-Walker background. 
We discuss a number of issues that arise when regularizing the theory with a momentum-space cutoff, and show explicitly how introducing non-covariant counter-terms allows to obtain covariant results for the renormalized vacuum energy-momentum tensor. We clarify some confusion in the literature
concerning the equation of state of vacuum fluctuations.
Further, we point out  that the general structure of 
the  effective action
becomes richer if the theory contains a scalar field $\phi$ with mass $m$ smaller than the Hubble parameter $H(t)$. Such an ultra-light particle cannot be integrated out completely to get the effective action. Apart from the volume term and the Einstein-Hilbert term, that are reabsorbed into renormalizations of the cosmological constant and Newton's constant, the effective action in general also has a term proportional to  $F(\phi)R$, for some function $F(\phi)$. As a result, vacuum fluctuations of ultra-light scalar fields naturally lead to models where the dark energy density has the form $\rho_{\rm DE}(t)=\rho_X(t)+\rho_Z(t)$, 
where $\rho_X$ is the component that accelerates the Hubble expansion at late times and $\rho_Z(t)$ is an extra contribution proportional to $H^2(t)$. We perform a detailed comparison of such models  with  CMB, SNIa and BAO data.
\end{abstract}


\maketitle



\section{Introduction}

The computation of the expectation value of the energy-momentum tensor in curved space, and particularly in a Friedmann-Robertson-Walker (FRW) background,
is a classic and much studied problem that has important implications  for understanding the origin of dark energy. 
The renormalization of 
$\cav\Tmn\vac$ in a FRW background can be performed  either with  covariant regularization schemes, such as dimensional regularization or point-splitting~\cite{Birrell:1982ix}, or using a cutoff $\lc$ in momentum space \cite{Parker:1974qw,Fulling:1974pu,Fulling:1974zr}. The former option has the advantage of preserving  general covariance explicitly, while a momentum-space cutoff can be more intuitive, for instance when developing naturalness arguments. 
The two schemes should give the same results for the renormalized quantities. 
However, the explicit breaking of general covariance by a momentum-space cutoff leads to some technical subtleties, that apparently have given rise to some confusion  in both the old and the recent 
literature. For instance it was claimed that regularizing a theory with a sharp momentum cutoff $\lc$ in a FRW background leads to pathologies~\cite{Fulling:1974zr} connected to the fact that the resulting vacuum energy density and pressure satisfy an equation of state (EOS) that is not consistent with general covariance \cite{Peebles:2002gy}. 
It was also claimed that energy-momentum conservation is violated by the flow of momentum modes across the cutoff, induced by the cosmological 
red-shift~\cite{KeskiVakkuri:2003vj,Mangano:2010hw,Sloth:2010ti}.
There even appears to be some confusion as to what general covariance implies for 
$\cav\Tmn\vac$.  For a constant term such as the $\lc^4$ divergence,   general covariance implies  that its contribution to $\cav\Tmn\vac$ must be proportional to the metric $\gmn$, so its EOS parameter is $w=-1$. However, 
contrary to some  statements in the literature 
(e.g.\ \cite{Padmanabhan:2004qc,Sloth:2010ti})
this is not true for a time-dependent term such as the quadratic divergence $H^2(t)\lc^2$, where $H(t)$ is the Hubble parameter. We will show that general covariance implies that the contribution of this term to $\cav\Tmn\vac$ must be proportional to the Einstein tensor $\Gmn$. The corresponding EOS parameter is time-dependent and equal to the total EOS parameter of the background,
$w_{\rm tot}(t)=p_{\rm tot}(t)/\rho_{\rm tot}(t)$, rather than being $w=-1$.

The aim of this paper is two-fold. First, we will  
clarify the above technical issues. When using a non-covariant regularization scheme one must also include non-covariant counter-terms. We will see explicitly how introducing the appropriate non-covariant counter-terms allows us to obtain fully covariant results for the renormalized quantities. 
As a consequence, no pathologies arise even when employing a sharp momentum cutoff. The renormalized energy density and pressure satisfy the standard energy-momentum conservation. The EOS parameters associated to the terms arising from both the quartic and the quadratic divergences are dictated by general covariance and fully agree with the results obtained within covariant regularization schemes. 
We will also discuss a possible alternative to the standard renormalization of vacuum energy, in which the subtraction of the flat-space contribution from the vacuum energy computed in FRW is performed already at the purely classical level, invoking arguments from the ADM formalism of General Relativity, as well as from AdS/CFT correspondence, developing an argument that we presented in 
\cite{Maggiore:2010wr,Maggiore:2011hw}.

After having shown that no new physics appear merely as a consequence of employing a non-covariant regularization scheme, 
we will turn to 
the second aim of this paper, which is to show that, in contrast,  the general structure of 
$\cav\Tmn\vac$ changes in an interesting manner if the theory contains
a scalar field $\phi$ whose mass $m$ is smaller than the Hubble parameter $H(t)$ at the time of interest. In particular, this applies to an ultra-light scalar field with $m<H_0$. In this case the vacuum expectation value (VEV) of the energy-momentum tensor $\Tmn$ acquires  extra terms, compared to the standard case where all matter fields are  very massive compared to $H(t)$. In general the VEV of $\Tmn$ can be obtained by functional differentiation,
\be\label{TmnSeff}
\cav\Tmn\vac=-\frac{2}{\sqrt{-g}}\frac{\d S_{\rm eff}}{\d\gMN}\, ,
\ee
where $S_{\rm eff}$ is
the effective action for gravity, which is
obtained by treating the metric $\gmn$ as a classical background and integrating out the massive matter degrees of freedom
(see e.g.\ refs.~\cite{Barvinsky:1985an,Buchbinder:1992rb,Mukhanov:2007zz,Shapiro:2008sf}).  One can perform the calculation of the effective action using manifestly covariant regularization schemes such as dimensional regularization or point-splitting. The effective action is then explicitly generally covariant and depends only on structures such  as the volume term $\int d^4x\sqrt{-g}$,  the Einstein-Hilbert term 
$\int d^4x\sqrt{-g} R$, and higher-derivative terms such as 
$\int d^4x\sqrt{-g} \,\Rmn R^{\mu\nu}$. The volume term is reabsorbed into a renormalization of the cosmological constant (giving rise to the cosmological constant naturalness problem) while the term in the effective action proportional to the Einstein-Hilbert term  renormalizes Newton's constant. Higher-derivative terms such as $\Rmn
R^{\mu\nu}$ can be seen as genuinely new effects due to vacuum fluctuations, but they can only be relevant when the  curvature radius is of order of the Planck length. In particular, their impact  is utterly negligible in the recent cosmological epoch.

The basic idea we develop in this paper is that if a scalar field $\phi$ with mass  $m<H_0$ is present it cannot be integrated out completely, since it is not massive with respect to the energy scale of interest in cosmology, $H(t)$. Thus, after integrating over the massive degrees of freedom, the effective action will be a functional of both the metric $\gmn$ and the field $\phi$ (or, more precisely, of its low-frequency modes, that we will denote collectively as $\chi$). This allows for a richer structure of the effective action and therefore of the corresponding $\cav\Tmn\vac$. In particular, the effective action will generically include a term $\propto \int d^4x\sqrt{-g} \, f(\chi)R$, with some function $f(\chi)$. Even at low curvatures such a term is not suppressed with respect to the Einstein-Hilbert term, and can have interesting phenomenological consequences. We will see that this indeed leads to an explicit realization of a  model of interacting vacuum fluctuations recently proposed in 
refs.~\cite{Maggiore:2010wr,Maggiore:2011hw}, where the dark energy tracks the total energy density of the Universe. 

The paper is organized as follows. In section \ref{sect:renorm} we discuss the renormalization of vacuum energy in FRW using a momentum-space cutoff.
In section~\ref{sect:chi} we discuss the effect  of an ultra-light scalar field 
on $\cav\Tmn\vac$. We then construct a cosmological model based on these considerations and we  analyze its phenomenological consequences.

We use signature ($-$+++) and  natural units where $\hbar=c=1$, so $G=\mpl^{-2}$, where $\mpl$ is the Planck mass. When we  specialize our results to FRW,  we work in a spatially flat FRW metric with  cosmic time $t$, scale factor $a(t)$ and Hubble parameter $H(t)=(1/a)da/dt=\dot{a}/a$. Today, the Hubble parameter and the critical density take the values $H_0$ and $\rho_0=3H_0^2/(8\pi G)$, respectively.

\section{Vacuum energy and renormalization in FRW space-time}\label{sect:renorm}

\subsection{Renormalization and general covariance with a momentum-space cutoff}\label{sect:rencovar}

To illustrate the kind of problems that arise when computing $\cav\Tmn\vac$ with a cutoff $\lc$ over momenta, let us begin with the $\lc^4$ divergence in flat space. Considering for instance a single massless real scalar field, a straightforward calculation gives
\be\label{rbare}
\rho_{\rm bare}(\lc)\equiv \cav T_{00}\vac =\frac{\lc^4}{16\pi^2}\, ,
\ee
where the subscript ``bare" emphasizes that this is still a bare, cutoff-dependent quantity, rather than the physical, renormalized energy density.
For the pressure, which is given by  $p
=(1/3)\sum_i \cav T_{ii}\vac$, an analogous computation gives \cite{Fulling:1974zr,Akhmedov:2002ts}
$p_{\rm bare}(\lc)=\lc^4/(48\pi^2)$, and therefore 
\be\label{pbare}
p_{\rm bare}(\lc)=\frac{1}{3}\,\rho_{\rm bare}(\lc)\, . 
\ee
This result has sometimes lead to the interpretation that the EOS of vacuum fluctuation is $p=w\rho$ with $w=1/3$ (see e.g.\ footnote~19 of ref.~\cite{Peebles:2002gy}). In contrast, 
using a scheme such as dimensional regularization  that preserves Lorentz invariance, in flat space one automatically gets $\cav\Tmn\vac\propto\emn$ and therefore $p=-\rho$.

The resolution of this apparent discrepancy is that the energy density and pressure 
given in \eqs{rbare}{pbare} are bare quantities rather then renormalized ones and, in a regularization scheme that breaks Lorentz invariance, the EOS parameter $w_{\rm bare}\equiv
p_{\rm bare}/\rho_{\rm bare}$ is not the same as the physical EOS parameter for the renormalized quantities, $w_{\rm ren}\equiv p_{\rm ren}/\rho_{\rm ren}$.
Indeed, regularizing the theory with a cutoff over momenta breaks the Lorentz symmetry of Minkowski space, since the notion of maximum spatial momentum is not invariant under boosts. This means that in this scheme the renormalization procedure also involves counter-terms that are not Lorentz invariant. To give an explicit example of such a counter-term it is convenient to work with a generic metric $\gmn$  at first, before setting $\gmn=\emn$ at the end. This allows us to derive the vacuum energy-momentum tensor from $\cav\Tmn\vac=-(2/\sqrt{-g})\, \d S/\d\gMN$, a procedure that facilitates the extension to curved space-time. The standard Lorentz-invariant counter-term in the action is proportional to the volume term 
\be\label{volterm}
S_A =
A(\lc) \int d^4x\sqrt{-g}\, , 
\ee
with $A(\lc)$ a  suitably chosen function of the cutoff.
An example of a Lorentz-breaking counter-term is 
\be\label{volterm2}
S_B =
B(\lc) \int d^4x\sqrt{-g}\,g^{00}\, .
\ee
Using $\d(\sqrt{-g}) =
-(1/2)\sqrt{-g}\, \gmn\d\gMN$ we see that
the contribution to $\cav\Tmn\vac$ from $S_A$ is 
\be
-\frac{2}{\sqrt{-g}}\, \frac{\d S_A}{\d\gMN} =A(\lc)\gmn\, ,
\ee
while the contribution from $S_B$ is
\be
-\frac{2}{\sqrt{-g}}\, \frac{\d S_B}{\d\gMN} =B(\lc)
(g^{00}\gmn -2\d_{\mu}^0\d_{\nu}^0)\, .
\ee
Evaluating them on the flat metric $\gmn=\emn$ and using the flat-space expression
$\Tmn=(\rho, p, p, p)$ (or, more generally, using the
FRW metric $\gmn= (-1,a^2,a^2,a^2)$ and $\Tmn=(\rho, a^2p, a^2p, a^2p)$), we see that $S_A$ produces a counter-term for the energy density given by $\rho_A=-A(\lc)$ and a counter-term  $p_A=-\rho_A=A(\lc)$ for the pressure. In contrast $S_B$ produces the counter-terms $\rho_B=+p_B=-B(\lc)$.

For illustration we consider again a minimally coupled massless scalar field where $\rho_{\rm bare}(\lc)=\lc^4/(16\pi^2)$ and $p_{\rm bare}(\lc)=\lc^4/(48\pi^2)$. After adding these counter-terms to the bare $\lc^4$ divergences we get
\bees
\rho_{\rm ren}&=&\rho_{\rm bare}(\lc)+\rho_{\rm count}(\lc)\nn\\
&=&
\frac{\lc^4}{16\pi^2} - A(\lc)-B(\lc)\, ,\\
p_{\rm ren}&=&p_{\rm bare}(\lc)+p_{\rm count}(\lc)\nn\\
&=&
\frac{\lc^4}{48\pi^2} + A(\lc)-B(\lc)\, .
\ees
We can now choose $A(\lc)$ and $B(\lc)$ such that  $A+B=\lc^4/(16\pi^2) -
\rho_{\rm finite}$ (where $\rho_{\rm finite}$ is a finite part)
 and $A-B=-\lc^4/(48\pi^2)-\rho_{\rm finite}$, i.e.\ $A(\lc)=\lc^4/(48\pi^2)-\rho_{\rm finite}$ and $B(\lc)=2\lc^4/(48\pi^2)$. In this way  we get
$\rho_{\rm ren}=\rho_{\rm finite}$ and $p_{\rm ren}=-\rho_{\rm finite}$, and therefore the renormalized quantities satisfy $p_{\rm ren}=-\rho_{\rm ren}$.
Thus, Lorentz invariance can be restored for the renormalized quantities despite the fact that it was broken by the regularization scheme in the first place.

Having understood this simple point in the flat-space case, we can apply a similar strategy in curved space, except that now the guiding symmetry principle is no longer Lorentz invariance, but rather general covariance. In curved space one can again use regularization schemes, such as dimensional regularization or point-splitting, that preserve general covariance explicitly, and in this case one finds a fully covariant result already at the level of the bare quantities. In particular, $\cav\Tmn\vac$ can be written as in \eq{TmnSeff}, where $S_{\rm eff}$ is made of coordinate-invariant quantities. For instance, for a minimally coupled scalar field in $d=4-\eps$ dimensions, the UV divergent part of the effective action has the form (see e.g.\ eqs.\ (6.44)-(6.49)
of ref.~\cite{Birrell:1982ix})
\bees\label{Seffcov}
S_{\rm eff}&=&\int d^4x\sqrt{-g}\[ c_1(\eps) +c_2(\eps) R \right.\\
&&\left. +c_3(\eps)
\( R_{\mu\nu\rho\sigma}R^{\mu\nu\rho\sigma} -R_{\mu\nu}R^{\mu\nu}
-6\Box R\)\]\, ,\nn
\ees
where $c_1(\eps), c_2(\eps), c_3(\eps)$ contain a term diverging as $1/\eps$ and a finite part.
It is interesting to compare this with the structure of divergences obtained using a cutoff $\lc$ over momenta, and specializing for simplicity directly to a FRW background. In this case, considering for definiteness a massless and minimally coupled real scalar field, we get
$(\cav T^{\mu}_{\nu}\vac)_{\rm bare} = {\rm diag}
(-\rho_{\rm bare}, p_{\rm bare},p_{\rm bare},p_{\rm bare})$, where \cite{Fulling:1974zr}
\bees  
\rho_{\rm bare}(\lc)
&=&\frac{\lc^4}{16\pi^2}+\frac{H^2(t)\lc^2}{16\pi^2}+{\cal O}(H^4\ln\lc)\, ,\label{rhovac}\\
p_{\rm bare}(\lc)&=&\frac{\lc^4}{48\pi^2}+c_1\frac{H^2(t)\lc^2}{16\pi^2}
+{\cal O}(H^4\ln\lc)\, .\label{pvac}
\ees
The coefficient $c_1$ depends on the cosmological epoch: $c_1=-1/3$ during De~Sitter (DS), $c_1=+1$ during radiation dominance (RD) and $c_1=2/3$ during matter dominance (MD) 
\cite{Maggiore:2010wr}.

The $\lc^4$ divergence is of course the same as in flat space. Again, it cannot be renormalized simply using the covariant counter-term (\ref{volterm}), since the coefficient of the $\lc^4$ term in the pressure is not equal to minus the coefficient 
of the $\lc^4$ term in the energy density. Rather, one must also introduce the counter-term (\ref{volterm2}) and adjust these two counter-terms so that the corresponding renormalized energy density and pressure satisfy $p_{\rm ren}=-\rho_{\rm ren}$. An equivalent way to 
phrase this point is to
observe that the $\lc^4$ terms in $\rho_{\rm bare}(\lc)$ and $p_{\rm bare}(\lc)$ do not satisfy the continuity equation, i.e.
\be\label{contbare}
\dot{\rho}_{\rm bare}+3H (\rho_{\rm bare}+p_{\rm bare})\neq 0\, ,
\ee
since for these quartically divergent terms $\dot{\rho}_{\rm bare}=0$ but $\rho_{\rm bare}+p_{\rm bare}\neq 0$. This is just a consequence of the fact that this regularization scheme breaks general covariance. Therefore energy-momentum conservation, which is a consequence of general covariance, does not hold for the bare quantities. There is nothing wrong or unphysical about this, as long as we choose the counter-terms such that the {\em renormalized} energy density and pressure satisfy energy-momentum conservation, which they do, since we saw that we can renormalize the quartic divergence so that, for this term, $p_{\rm ren}=-\rho_{\rm ren}$.

Similar but somewhat more interesting considerations can be made for the quadratically divergent term. First of all, let us see what  the EOS of the term $\propto H^2(t)\lc^2$  should be, in order to satisfy energy-momentum conservation. Writing $\rho(t)=c H^2(t)$,
where $c$ is a constant, 
$p(t)=w(t)cH^2(t)$,  and plugging this into the conservation equation
$\dot{\rho}+3H (\rho+p)= 0$,
 we get
\be\label{wzdH2da}
3[1+w(t)]=-\frac{1}{H^3}\frac{dH^2}{dt}=-\frac{a}{H^2}\, \frac{dH^2}{da}\, ,
\ee
Using the Friedmann equation $H^2(t)=(8\pi G/3)\sum_i\rho_i(t)$, where the index $i$ runs through all contributions to the energy density of the Universe (matter, radiation, dark energy, etc.) and using $\rho_i(t)\propto a^{-3(1+w_i)}$, we get
\be\label{wwtot}
w(t)= \frac{\sum_i w_i\rho_i(t)}{\sum_i\rho_i(t)}\equiv w_{\rm tot}(t)\, .
\ee
This is simply the total EOS parameter $w_{\rm tot}(t)=p_{\rm tot}/\rho_{\rm tot}$ that tracks the EOS parameter of the dominant component, so it evolves from $w_{\rm tot}(t)\simeq 1/3$ during RD to $w_{\rm tot}(t)\simeq 0$ during MD, and finally $w_{\rm tot}(t)\simeq -1$ if there is a dark energy dominated phase with $w_{\rm DE}\simeq-1$.

We can now see that the terms $\propto H^2(t)\lc^2$ in \eqs{rhovac}{pvac} do not satisfy energy-momentum conservation, since the bare EOS parameter $c_1$ is not equal to $w_{\rm tot}(t)$. Rather, 
the values $c_1=\{+1,+2/3,-1/3\}$ computed assuming  a pure  RD, MD and DS epoch, respectively, are reproduced by 
\be\label{c1w}
c_1(t)=w_{\rm tot}(t)+2/3\, .
\ee
The situation is completely analogous to that discussed for the $\lc^4$ divergence. If the quadratically divergent terms in the bare energy density and pressure were satisfying energy-momentum conservation we could renormalize them with a covariant counter-term, which in this case would  simply be the Einstein-Hilbert term 
$\int d^4x\sqrt{-g}\, R$, so they would just be reabsorbed into a renormalization of Newton's constant. The fact that the quadratically divergent terms in the  bare energy density and pressure do not satisfy energy-momentum conservation means that we must also introduce  a non-covariant counter-term, chosen so that the contribution proportional to $H^2(t)$ in the renormalized energy density and pressure satisfy 
$\rho_{\rm ren}(t)={\rm const.}\times H^2(t)$ and $p_{\rm ren}(t)=w_{\rm tot}(t)\rho_{\rm ren}(t)$. 
We explicitly construct the required non-covariant counter-term in appendix~\ref{app:noncov}.

To conclude this section, let us observe that a dark energy term $\propto H^2(t)\mpl^2$ can be obtained rather generally, for instance from holographic considerations. In this context it is sometimes stated that, if the regularization
of the divergent expressions is handled in a Lorentz-invariant manner, as with dimensional regularization,  the corresponding VEV of the energy-momentum tensor will have the form
$\cav\Tmn\vac\propto\gmn$, i.e.\ $w=-1$~\cite{Padmanabhan:2004qc}.
The discussion  presented above shows that this is not correct. First of all, Lorentz invariance is irrelevant here as it is not a symmetry of the FRW space-time. Rather, the relevant symmetry is general covariance. As we have seen above, for 
the term $\propto H^2(t)$ in the energy density, general covariance fixes the EOS parameter to $w=w_{\rm tot}(t)$ rather than to $w=-1$. The two results coincide only in a De~Sitter epoch. 

An equivalent way of understanding this point is to observe that a term $\propto H^2(t)$ in $\cav\Tmn\vac$ is obtained by taking the functional derivative of the Einstein-Hilbert term in the effective action (and not of the volume term). Therefore its covariant form is proportional to the Einstein tensor $G_{\mu\nu}$ rather than to the metric
$g_{\mu\nu}$ (see also the explicit covariant computation in \cite{Bilic:2011rj}). Again, if $H(t)$ is constant in time $G_{\mu\nu}\propto\gmn$ and the two results agree, otherwise they are different.
This is also the most direct way of deriving the EOS for the term $\propto H^2(t)$ in  $\cav\Tmn\vac$: using the fact that this contribution to $\cav\Tmn\vac$ is proportional to $G_{\mu\nu}$, together with Einstein equations $G_{\mu\nu}=(8\pi G) T^{\rm tot}_{\mu\nu}$  we  see that in the end this contribution to 
$\cav\Tmn\vac$  is proportional to $T^{\rm tot}_{\mu\nu}$.  So, in FRW, the term proportional to $H^2(t)$ in the vacuum energy density and pressure satisfy $p_{\rm vac}/\rho_{\rm vac}=p_{\rm tot}/\rho_{\rm tot}\equiv w_{\rm tot}(t)$.

\subsection{Comparison with previous work}

The fact that the terms $\propto H^2(t)\lc^2$  in $\rho_{\rm bare}(\lc)$ and $p_{\rm bare}(\lc)$
do not respect energy-momentum conservation appears to have generated some recurrent confusion in the literature. Renormalization of the energy-momentum tensor with a momentum-space cutoff in FRW was  discussed long ago in a  classic paper by 
Fulling and Parker \cite{Fulling:1974zr}. However, we believe that our discussion above sheds light on some aspects of ref.~\cite{Fulling:1974zr}. In  that paper, the authors considered 
the $(00)$ and the $(0i)$ components of the Einstein equations in a FRW space-time, 
both
written  in terms of the bare quantities\footnote{Compared to eq.~(5.4) of ref.~\cite{Fulling:1974zr},
for notational simplicity we limit to the spatially flat case and we do not explicitly write the term which is related to the renormalization of terms quadratic in the Riemann tensor
, denoted $H_{\mu\nu}$ in  ref.~\cite{Fulling:1974zr}.}
\bees
\(\frac{\dot{a}}{a}\)^2-\frac{1}{3}\Lambda_0&=&\frac{8\pi G_0}{3}\rho_{\rm bare}\, ,\label{eq00bare}\\
2\frac{\ddot{a}}{a}+\(\frac{\dot{a}}{a}\)^2-\Lambda_0 &=&-8\pi G_0 p_{\rm bare}\, ,\label{eqiibare}
\ees
where $\Lambda_0$ is the bare cosmological constant, not to be confused with the cutoff $\lc$, and $G_0$ the bare Newton's constant. Regularizing the theory with a sharp cutoff in momentum space, and restricting  for simplicity to the contribution of a  minimally-coupled massless scalar field in De~Sitter space, they find that both $\rho_{\rm bare}$ and $p_{\rm bare}$ have a quartic as well as a quadratic divergence,  given by
\eqs{rhovac}{pvac}, with $c_1=-1/3$ for De~Sitter. Including also the finite parts
$\rho_{\rm finite}$ and $p_{\rm finite}$ we therefore have
\bees
\rho_{\rm bare}(\lc)
&=&\frac{\lc^4}{16\pi^2}+\frac{H^2\lc^2}{16\pi^2}+{\cal O}\(\ln \lc\) +\rho_{\rm finite}
\, ,\label{rhovac1}\\
p_{\rm bare}(\lc)
&=& \frac{\lc^4}{48\pi^2}-\frac{H^2\lc^2}{48\pi^2}
+{\cal O}\(\ln \lc\) +p_{\rm finite}\, .\label{pphikd3k2}
\ees
Thus \eq{eq00bare} can be rewritten as 
\be\label{renGFP}
\(\frac{\dot{a}}{a}\)^2-\frac{1}{3}\(
\frac{\Lambda_0+\frac{G_0\lc^4}{2\pi}}{1-\frac{G_0\lc^2}{6\pi}}\)=
\frac{8\pi }{3}\, \(\frac{G_0}{1-\frac{G_0\lc^2}{6\pi}}\)\,\rho_{\rm finite}\, 
\ee
which shows that, at least in this equation, the quartic divergence is reabsorbed into a renormalization of the cosmological constant and the quadratic divergence into a renormalization of Newton's 
constant.\footnote{The logarithmically divergent terms, that we have not written explicitly, are reabsorbed into the  renormalization of higher derivative operators such as $R_{\mu\nu} R^{\mu\nu}$
\cite{Fulling:1974zr}.} 

The situation is however different in \eq{eqiibare} since, if we now insert the explicit expression 
of $p_{\rm bare}$, we see 
that in this equation neither the quartic nor the quadratic divergence can be reabsorbed
into  the  above renormalization of $\Lambda_0$ and $G_0$. The authors of ref.~\cite{Fulling:1974zr} propose the following solution. They observe that \eq{eqiibare} can be obtained by taking a time derivative of \eq{eq00bare}, upon use of the conservation equation $\dot{\rho}_{\rm bare}+3H (\rho_{\rm bare}+p_{\rm bare})=0$.
However, as we saw above,  $\rho_{\rm bare}$ and $p_{\rm bare}$ given in
\eqs{rhovac1}{pphikd3k2}, do not satisfy this conservation equation, neither for the quartically divergent terms nor for the quadratically divergent terms.
They then discard the results (\ref{rhovac1}) and (\ref{pphikd3k2}) arguing that a sharp cutoff in momentum space is not a correct regularization, and that one should rather evaluate $\rho_{\rm bare}$ using a smooth cutoff function, i.e.\ replacing
\be\label{FP prescr}
\int_0^{a\lc}dk \ra \int_0^{\infty}dk\, f(k/a\lc)\, ,
\ee
where $f(k/a\lc)$ is a smoothed version of the theta function $\theta(1-k/a\lc)$. (Here $k$ denotes the comoving momentum, while in terms of the physical momentum, $k_{\rm phys}=k/a(t)$, the cutoff is at $\lc$.) Further, they argue that $p_{\rm bare}$ should be {\em defined} by 
$\dot{\rho}_{\rm bare}+3H (\rho_{\rm bare}+p_{\rm bare})\equiv 0$.\footnote{Actually, they write $p_{\rm bare}={\cal D}\rho_{\rm bare}\equiv (3a^2\dot{a})^{-1}
d(a^3\rho_{\rm bare})/dt$, see eqs.~(5.5) and (5.7) of ref.~\cite{Fulling:1974zr}, which is the same as $\dot{\rho}_{\rm bare}+3H (\rho_{\rm bare}+p_{\rm bare})=0$. The role of the smoothing function $f(k/a\lc)$ is that the time derivative in the operator ${\cal D}$ acts also on it, because of the factor $a(t)$ in the argument of $f$, generating a term proportional to $f'$, without which it would not be possible to impose 
\eq{contbare} to be satisfied.} In summary, according to ref.~\cite{Fulling:1974zr} the bare energy density should be computed as an integral over modes with a cutoff given by a smoothed theta function,  rather than with  a sharp cutoff,
while the bare pressure, instead of being given by the corresponding integral over modes with a smoothed theta function, should be {\em defined} by the conservation equation. 

Our discussion in the previous subsection however allows us to get a better understanding of this issue.
First of all,  there is nothing wrong with using a sharp momentum-space cutoff, and physical results should not depend on the choice of the regulator. 
The fact that it is impossible to reabsorb the divergences in \eqs{eq00bare}{eqiibare} into a renormalization of Newton's constant (by means of a counter-term proportional to the Einstein-Hilbert term) is not a sign of a pathology. Instead, it is a direct consequence of the need for non-covariant counter-terms to restore the symmetry after regularizing with a scheme that breaks general covariance. The Fulling-Parker prescription, which consists of using \eq{FP prescr} and defining $p_{\rm bare}(\lc)$ from $\dot{\rho}_{\rm bare}+3H (\rho_{\rm bare}+p_{\rm bare})=0$, can be seen as the definition of a regularization scheme that tries to enforce energy-momentum conservation by hand at the level of the bare quantities even in the presence of a momentum-space regularization. In this sense it is a legitimate, but certainly unnecessary (and somewhat akward) choice of regularization.\footnote{It should also be observed that general covariance is not fully recovered at the level of bare quantities even with this prescription. 
This clarifies another 
point of ref.~\cite{Fulling:1974zr} that the authors themselves considered disturbing, namely the fact that
some convergent terms in the expression for the bare energy density of order $m^2T^{-4}$, where $m$ is the mass of the scalar field and $T$ the parameter of adiabatic regularization, do not form the $(00)$ component of any divergenceless tensor, see the discussion at the bottom of page~186 of \cite{Fulling:1974zr}. Once again, the point is that if the regularization scheme breaks general covariance, the bare quantities will not respect it either. On the other hand, a generally covariant result for the renormalized quantities can  be obtained upon use of non-invariant counter-terms.}

A certain confusion seems to exist even in the more recent literature.  
When using a momentum-space cutoff it is not always appreciated that, even if the bare quantities do not satisfy energy-momentum conservation,
\be
\dot{\rho}_{\rm bare}+3H (\rho_{\rm bare}+p_{\rm bare})\neq 0\, , 
\ee
it is still possible to renormalize the theory such that the renormalized quantities satisfy it,
\be
\dot{\rho}_{\rm ren}+3H (\rho_{\rm ren}+p_{\rm ren})= 0\, . 
\ee
Equivalently,  when the regularization scheme breaks general covariance one needs to introduce non-covariant counter-terms and as a consequence the bare EOS parameter $w_{\rm bare}\equiv p_{\rm bare}/\rho_{\rm bare}$ is different from
the physical EOS parameter $w\equiv p_{\rm ren}/\rho_{\rm ren}$.
For instance, in ref.~\cite{Mangano:2010hw}  the fact that the bare vacuum energy-momentum tensor does not satisfy the conservation equation leads the author to propose a modification of the Einstein equations of the form $G_{\mu\nu} +\Sigma_{\mu\nu}=8\pi G\Tmn$, where $\Sigma_{\mu\nu}$ is a tensor that satisfies $\n_{\mu}\Sigma^{\mu\nu}
=8\pi G\n_{\mu}\TMN\neq 0$. However, once again, there is no special physical meaning in the fact that the {\em bare} energy-momentum tensor is not conserved. What appears on the right-hand side of the Einstein equations really is the conserved, physical energy-momentum tensor constructed from the renormalized vacuum energy density and pressure.

Similarly, in
ref.~\cite{Sloth:2010ti} the EOS of the term $H^2(t)\lc^2$ was investigated and it was claimed that 
the EOS parameter for vacuum fluctuations is $w=-1+\lc^2/(9\pi\mpl^2)$. However, this is a statement about the ratio $p_{\rm bare}(\lc)/\rho_{\rm bare}(\lc)$ and not about the ratio of the corresponding renormalized quantities.\footnote{Futhermore, the derivation 
in ref.~\cite{Sloth:2010ti} assumed that the EOS parameter consistent with general covariance is $w=-1$ while, as we saw above, for the term $\propto H^2(t)\lc^2$ general covariance gives $w=w_{\rm tot}(t)$.} Similar comments hold for the discussion of the EOS in \cite{Bilic:2011zm}.

The same criticism applies if one takes an effective field theory point of view rather than working in the context of full renormalization theory.  Let us
consider the theory which has been used to derive the result $\rho_{\rm bare}(\lc)=\lc^4/(16\pi^2)$ 
as the low-energy limit of a more fundamental theory that takes over at a scale $M$.  The total physical vacuum energy density will be given by the integral over the modes with $k<M$, that simply gives  $M^4/(16\pi^2)$, plus the contribution from the modes with $k>M$ which can only be computed using the UV completion of the theory and will be of order $M^4$, but comes with a numerical coefficient that  cannot be computed with only the low-energy theory at hand. Thus, the exact coefficients of $\lc^4$ in the energy density and the pressure are unpredictable within an effective field theory and, therefore, the physical EOS parameter cannot be computed either. Thus,
either one fixes the EOS parameter by means of symmetry principles, i.e.\ by demanding Lorentz invariance for the $\lc^4$ term or general covariance for the $H^2\lc^2$ term, or one simply has no handle on it. In any case, the EOS parameter derived from the ratio 
$p_{\rm bare}(\lc)/\rho_{\rm bare}(\lc)$ has no physical meaning.

\subsection{An alternative to cosmological constant renormalization: subtractions in classical GR}\label{sect:alt}

The renormalization of the $\lc^4$ divergence gives rise to the infamous cosmological constant naturalness problem: the divergence in the bare vacuum energy, $\rho_{\rm bare}(\lc)\equiv\cav T_{00}\vac$, is canceled by a counter-term
$\rho_{\rm count}(\lc)$ that corresponds to a counter-term proportional to  
$\int d^4x\,\sqrt{-g}$ in the action. The physical, renormalized vacuum energy is then $\rho_{\rm ren}=\rho_{\rm bare}(\lc)+\rho_{\rm count}(\lc)$. The naturalness problem arises from the fact that $\rho_{\rm count}(\lc)$ must cancel the $\lc^4$ divergence in $\rho_{\rm bare}(\lc)$, leaving a finite part that is exceedingly small compared to $\lc^4$, for all values of the cutoff larger than, say, the TeV scale, where quantum field theory is well tested.
 
At first sight the necessity of an extremely fine-tuned cancellation between $\rho_{\rm bare}(\lc)$ and
$\rho_{\rm count}(\lc)$ can be disturbing.
It should however be observed that neither the bare term  nor the counter-term  have any physical
meaning and only their sum is physical, so this fine tuning is  different
from an implausible cancellation between observable quantities. The same kind
of cancellation appears for instance in the Casimir effect, and this might provide
a  first hint of what could be the correct treatment of vacuum energies in cosmology. 
In the Casimir effect
one computes the vacuum energy density of a field in a finite volume
(e.g.\ the electromagnetic field between two parallel conducting plates at a distance $L$)
and subtracts from it the vacuum energy density computed in an infinite volume. If we regularize with a cutoff $\lc$ over momenta both terms diverge as $\lc^4$, but their difference is finite and depends only on the macroscopic scale $L$. Thus the quantity of interest 
is the difference between the vacuum energy in the given geometry and the vacuum energy in  a reference geometry, which is just  flat space-time in an infinite volume. This  might  suggest that, to obtain the physical effect of the vacuum energy density on the expansion of the Universe, one should analogously compute the vacuum energy density in a FRW space-time and subtract from it the value computed in a reference space-time, which could be naturally taken as Minkowski space. This procedure  leads to a sort of  ``cosmological Casimir effect". 

An immediate objection to this analogy could be that 
in special relativity the zero of the energy can be chosen arbitrarily and only energy differences with respect to the ground state are relevant. In contrast, in GR we cannot choose the zero of the energy arbitrarily. One typically expects that ``every form of energy gravitates", such that the contribution of Minkowski space cannot simply be dropped.
While it is certainly true that in GR the choice of the zero of the energy is not arbitrary, the point that we wish to make here is that  the correct choice  can be a non-trivial issue (see also \cite{Maggiore:2010wr,Maggiore:2011hw}).
As a first example, consider the definition of energy in  GR for asymptotically flat metrics.
To carefully define the energy associated with a given  field configuration it is convenient to 
use the Hamiltonian formulation of GR, which goes back to the classic paper by 
Arnowitt, Deser and Misner (ADM)
\cite{Arnowitt:1962hi} (see e.g.\ the textbook \cite{pois04} for a very clear recent review).
In order to define the Hamiltonian of GR one must  first work in a finite three-dimensional volume, where it  takes the form
\be\label{HHH}
H_{\rm GR}=H_{\rm bulk}+H_{\rm boundary}\, .
\ee
The bulk term $H_{\rm bulk}$ is given by an integral over the three-dimensional finite spatial volume at fixed time, and 
$H_{\rm boundary}$ by an integral over its two-dimensional boundary. If we try to define the energy of a classical field configuration as the value of this Hamiltonian evaluated on the classical solution we encounter a problem: as a consequence of the invariance under diffeomorphisms, the volume term $H_{\rm bulk}$ vanishes when evaluated on any classical solution of the equations of motion. Then the whole energy comes from the boundary term. However, when the boundary term is evaluated on any asymptotically flat metric (including Minkowski) it diverges as the boundary is taken to infinity. The solution proposed by ADM is to subtract the same boundary term evaluated on Minkowski space-time. The resulting energy (or mass) is finite and provides the standard definition of mass in GR, known as the ADM mass.
For instance, when applied to the \Sch space-time, the ADM mass computed in this way turns out to be equal to the mass   $M$ that appears in the \Sch metric.
The ADM prescription can be summarized by saying that, in GR, the energy $E$ 
associated to a classical asymptotically-flat space-time with metric
$\gmn$ can be obtained from the Hamiltonian $H_{\rm GR}$ through
\be\label{EHH}
E=H_{\rm GR}[\gmn]-H_{\rm GR}[\emn]\, ,
\ee
where $\emn$ is the flat metric.  Even if the context in which this formula is valid, namely asymptotically flat space-times, is  different from the cosmological context that we are interested in here, \eq{EHH}  still  suffices to make the point that the intuitive understanding that GR requires  any form of energy to act as a source for the gravitational field is not really correct. \Eq{EHH} tells us that the energy associated to Minkowski space does not gravitate.

The ADM prescription has also been generalized to space-times that are not asymptotically flat, 
subtracting the contribution of some reference space-time whose boundary has the same induced metric as the geometry under consideration~\cite{Gibbons:1976ue,Brown:1992br,Brown:1994gs,Hawking:1995fd}. The issue has been studied in detail in particular for  asymptotically AdS space-times, where the  problem appears because the boundary term in the Hamiltonian diverges as the boundary used in its definition approaches the  boundary of AdS. In this case
 it has been shown that the appropriate subtraction can be
performed without even introducing a reference background, but  by adding some local counter-terms to the boundary action. These are given by  coordinate-invariant functionals of the intrinsic boundary geometry,  fixed by the requirement to obtain a finite energy as the boundary is taken to infinity \cite{Balasubramanian:1999re,Kraus:1999di}. 
This prescription is particularly appealing  since for AdS  the structure of the divergences of the boundary action  is such that they
can be  removed by adding a finite polynomial in the boundary curvature and its
derivatives. This is analogous to the fact that in QFT the UV divergences are removed by  adding counter-terms that are  polynomials of finite order in the fields. Indeed, in the context of the AdS/CFT correspondence, 
this way of removing divergences in the gravitational action on the AdS side corresponds to the renormalization of the UV divergences in the conformal quantum field theory that lives on the 
boundary \cite{Balasubramanian:1999re,Kraus:1999di,Lau:1999dp,Mann:1999pc,Emparan:1999pm}.
This subtraction procedure correctly reproduces the masses of various known space-times that are asymptotically AdS. For instance, in $2+1$ dimensions it correctly gives the mass
of the BTZ black hole (which is asymptotically ${\rm AdS}_3$) and also reproduces the transformation law and conformal anomaly of the stress tensor in the dual CFT. In $3+1$ dimensions it correctly reproduces the known value of the mass of the ${\rm AdS}_4$-\Sch solution and similar results can be obtained for asymptotically ${\rm AdS}_5$ solutions. Furthermore, while this subtraction procedure provides a zero mass for pure   ${\rm AdS}_4$ space-times, it gives a non-vanishing value for the mass of pure  ${\rm AdS}_5$, which exactly matches the Casimir energy of the dual ${\cal N}=4$ super Yang-Mills theory that lives on the global ${\rm AdS}_5$ boundary 
with topology $S^3\times R$~\cite{Balasubramanian:1999re}. 

What we learn from these examples  is that the intuitive argument that ``GR requires that any form of energy contributes to the gravitational field" is in fact  loose and generally incorrect. Rather, in all cases  an appropriate subtraction, formulated either as the subtraction of the contribution of a reference space-time, or in terms of local counter-terms, is required already at the classical level in order to obtain an energy-momentum tensor which is well-defined and reproduces the  known properties of the space-time under consideration.

From this vantage point it is rather natural to assume  that  the definition of energy associated to zero-point quantum fluctuations in a curved background should  not escape  this general rule.  As we have seen, regularizing the theory with a cutoff $\lc$ over comoving momenta, in FRW the bare energy density of zero-point fluctuations  takes the  form
\bees\label{c1c2}
[\rho_{\rm bare} (\lc)]_{\rm FRW} &=&[\rho_{\rm bare} (\lc)]_{\rm Mink} 
+{\cal O}\(H^2(t)\lc^2\)\nn\\
&&+{\cal O}\(H^4(t)\ln\lc\)\, .
\ees
Here $[\rho_{\rm bare} (\lc)]_{\rm Mink}$ is the bare vacuum energy density in Minkowski space and,  for  a field of mass $m$, its  general UV structure  is
\bees\label{bare}
[\rho_{\rm bare}(\lc)]_{\rm Mink}&=&{\cal O}(\lc^4)+{\cal O}(m^2\lc^2)\nn\\
&&+{\cal O}(m^4\ln\lc)
+\mbox{\rm finite part}
\, .
\ees
For example, for a real minimally coupled massless scalar field, $[\rho_{\rm bare} (\lc)]_{\rm Mink}=c_1\lc^4$ and
\be
[\rho_{\rm bare} (\lc)]_{\rm FRW} =c_1\lc^4+c_2H^2(t)\lc^2+{\cal O}\(H^4(t)\ln\lc\)\, ,
\ee
with $c_1=c_2=1/(16\pi^2)$.\footnote{Actually, the value of $c_2$ also depends on the choice of modes, and hence of vacuum state, for the field under consideration. The value $c_2=1/(16\pi^2)$ is obtained using the modes that correspond to the usual Bunch-Davies vacuum. The most general result is obtained by taking a combination of these modes with arbitrary Bogoliubov coefficients. This changes the numerical value of the constant $c_2$, but not the general structure of divergences displayed in \eq{c1c2}, see e.g.\ appendix~B of \cite{Maggiore:2010wr}.\label{foot:BD}}
The quartic divergence is the same as that found in flat Minkowski space. This  can be seen from the explicit computation and is of course more generally a consequence of the fact that the result must reduce to that obtained in flat space in the limit $H\ra 0$.

Our discussion of the definition of energy in GR  suggests the following alternative to the standard renormalization of the cosmological constant. 
Rather than eliminating the $\lc^4$ divergence with standard renormalization, which allows us to get rid of it at the price of the naturalness problem, 
another option is that we should first perform the appropriate subtraction, corresponding to the correct choice of the zero of the energy density. To understand what is the correct subtraction in FRW, we consider the Friedmann equation when on the right hand side of the Einstein equations we use as source
$T_{\mu\nu}+\cav\Tmn\vac$, 
where $T^{\mu}_{\nu} ={\rm diag}(-\rho,p,p,p)$ is the ordinary classical contribution of matter, radiation, etc,
and $\cav\Tmn\vac$ is the corresponding contribution of zero-point fluctuations. The Friedmann equation then reads
\be
H^2(t)=\frac{8\pi G}{3}\( \rho+\cav T_{00}\vac\)\,.
\ee
We fix the subtraction procedure  by requiring that  Minkowski space (i.e.\ $H(t)=0$) should be  a solution in the limit  $\rho\ra 0$. This implies that all terms in 
$[\rho_{\rm vac}]_{\rm FRW}$ that do not vanish when $H\ra 0$ must be subtracted, i.e.\ we must subtract the vacuum energy computed in Minkowski space. This eliminates the $\lc^4$ divergence (as well as other flat-space divergences such as terms
$\propto m^2\lc^2$ that appear for a massive field with mass $m$)
in a way that does not suffer from the naturalness problem. The remaining divergences are renormalized in the standard way. In particular, the term $\propto H^2(t)\lc^2$ in the energy density is reabsorbed into the renormalization of Newton's constant (plus  non-covariant 
counter-terms for renormalizing the pressure, as discussed in section~\ref{sect:rencovar}). The renormalization of these remaining divergences does not suffer from
any naturalness problem (and, in any case, they could not be affected by a constant shift in the zero-point of the energy).\footnote{Needless to say, none of the experiments that are usually mentioned as probes of vacuum fluctuations, such as the Casimir effect or the Lamb shift, are affected by such a choice of the zero point of the energy. These experiments, of course, do not probe directly the choice of the zero in the definition of the energy; rather they always probe energy differences, while the zero of the energy is defined, arbitrarily, by normal ordering the Hamiltonian.}

It is also interesting to observe that this procedure
provides a new perspective for interpreting the results of ref.~\cite{Brustein:2000hh}. There the authors show that in a theory with  a large number ${\cal N}$ of quantum fields,  because of the finite volume fluctuations of vacuum energy around its average value,
Minkowski space would be unstable to black hole formation  in regions of size ${\cal O}({\cal N}^{1/4}\lpl)$, where $\lpl$ is the Planck length. This is of course a paradoxical result,  and to escape this conclusion the authors of ref.~\cite{Brustein:2000hh} suggest
that the quantum gravity length-scale for ${\cal N}$ fields should be  of order 
${\cal N}^{1/4}\lpl$ instead of simply $\lpl$. In other words the UV cutoff should be parametrically smaller than the Planck mass, and of order ${\cal N}^{-1/4}\mpl$. If this was the case,
semiclassical gravity would break down at this scale, and the
calculation used to prove black hole formation would no longer be valid.
With our proposal, in contrast, the problem is eliminated simply because vacuum fluctuations in Minkowski space do not gravitate.

By means of this approach it is also possible to understand why phase transitions in the early Universe do not produce an exceedingly 
large vacuum energy density~\cite{Zhou:2011um}. Consider for instance a scalar field $\varphi$ undergoing spontaneous symmetry breaking (SSB), with the typical double-well potential 
\be
V(\varphi)=-\frac{1}{2}\mu^2\varphi^2+\frac{\lambda}{4} \varphi^4\, , 
\ee
with $\mu^2>0$. Computing the energy density in the FRW background, one finds the vacuum energy
associated to SSB to be
\be\label{rhoSSB}
\rho_{\rm SSB}= -\frac{\mu^4}{4\lambda} -\frac{\mu^2}{2\lambda} (\dot{H}+2H^2)\, .
\ee
The first term is the flat-space result that would typically give an unacceptably high value for the vacuum energy density. For instance, the electroweak transition would give $\rho_{\rm SSB}^{1/4}={\cal O} (100\, {\rm GeV})$. With our proposal, however, this Minkowski space term does not contribute to the gravitating energy density, and one remains with the second term in \eq{rhoSSB}, that is ${\cal O}(\mu^2H^2/\lambda)$ and therefore totally consistent with observations.

\section{Vacuum energy and ultra-light scalars}\label{sect:chi}

The quadratic divergence proportional to $H^2(t)\lc^2$ has the interesting property that, if we set the cutoff $\lc$ at a value of the order of the Planck mass $\mpl$, it becomes  of the order of the critical density of the Universe at time $t$, since
$\rho_c(t)=(3/8\pi) H^2(t)\mpl^2$. This is in sharp contrast with the quartic divergence $\lc^4$, that for $\lc\sim\mpl$ is more than 120 order of magnitudes larger than the observationally allowed value.
Thus, the quadratic divergence might  offer a hint for understanding the coincidence problem, namely why the vacuum energy density that drives the acceleration of the Universe is  of the order of the critical density today. In this naive form, however, this argument does not lead us very far, since the quadratic divergence is simply reabsorbed into a renormalization of Newton's constant and does not leave observable effects. Still, the fact that a term proportional to $H^2(t)$ in the vacuum energy density might naturally be of the order of  the critical density is quite intriguing and stimulates further investigation. In this section we will see that it is indeed possible to prevent the term $\propto H^2(t)$ of being reabsorbed into a renormalization of Newton's constant. However, to do so we are required to introduce an ultra-light degree of freedom into the theory.
In this paper, for definiteness, we will consider a scalar field. However, most of our considerations in this section go through as well if we introduce an ultra-light massive graviton instead. The role of the scalar field is then played by the helicity-0 component of the massive graviton.
Ultra-light scalars with $m\sim H_0$ where first discussed,
in the context of pseudo-Nambu-Goldstone bosons, in \cite{Frieman:1995pm}.
In ref.~\cite{Sahni:1998at} it was first suggested that quantum effects from ultra-light scalar fields could lead to late-time acceleration.

\subsection{Ultra-light scalars and the effective action}

We begin by investigating the most general structure of the renormalized VEV of the energy-momentum tensor. First of all, to avoid misunderstandings, let us emphasize that the answer to this question does not depend on the regularization scheme employed. This may affect the structure of the {\em bare} $\cav\Tmn\vac$, but not that of the renormalized one. In a theory without ultra-light scalar fields, a generally covariant regularization gives the most general structure of the bare $\cav\Tmn\vac$ as in \eqs{TmnSeff}{Seffcov}. 
In this case  even the counter-terms are generally covariant, so the renormalized $\cav\Tmn\vac$ can in general have a term proportional to $\gmn$ from the variation of the volume term (which is  set to zero  if one adopts  the proposal in 
section~\ref{sect:alt}, or otherwise reabsorbed into a renormalization of the cosmological constant). Further it can have a term proportional to the Einstein tensor $G_{\mu\nu}$ from the variation of the Einstein-Hilbert term, reabsorbed into Newton's constant, and contributions from the variation of the terms quadratic in the Riemann tensor. On dimensional grounds the latter are suppressed by a factor $(E/\mpl)^2$ with respect to the Einstein-Hilbert term at the energy (or curvature) scale $E$, and can therefore only be relevant at Planckian energies.

As we have discussed, using  a momentum-space cutoff slightly complicates  the analysis as the explicit calculation of the bare
$\cav\Tmn\vac$ in such a scheme produces non-covariant contributions. These are compensated by introducing non-covariant counter-terms, chosen so to restore general covariance for the renormalized quantities. Thus, in the end the renormalized 
$\cav\Tmn\vac$  is independent of the regularization scheme, as it should. 
In the following, we will use a generally covariant regularization, so the general structure of the bare $\cav\Tmn\vac$ is the same as that of the renormalized 
$\cav\Tmn\vac$.

To understand if it is possible to obtain a more general result for $\cav\Tmn\vac$
it is useful to recall that the 
VEV of $\Tmn$ can be obtained by taking the functional derivative of the effective action with respect to the metric, see e.g.\ \cite{Mukhanov:2007zz,Shapiro:2008sf}.
The effective action
$S_{\rm eff}[\gmn]$ is derived from the matter action $S_m[\gmn,\psi]$ by integrating over the matter fields, here collectively denoted by $\psi$
\be\label{defSeff}
e^{i S_{\rm eff}[\gmn]}=\int{\cal D}\psi\, e^{iS_m[\gmn,\psi]}\, .
\ee
The matter energy-momentum tensor is defined as usual by
$T_{\mu\nu}=-(2/\sqrt{-g})\, \d S_m/\d\gMN$. In the path-integral formulation its vacuum expectation value is given by
\be
\cav\Tmn\vac = \frac{ \int{\cal D}\psi\, \Tmn  e^{iS_m[\gmn,\psi]}}
{\int{\cal D}\psi\, e^{iS_m[\gmn,\psi]}}\, ,
\ee
and can be rewritten as
\bees
\cav\Tmn\vac &=& \frac{1}{e^{i S_{\rm eff}[\gmn]}}
\(\frac{-2}{i\sqrt{-g}}\)\, \frac{\d }{\d\gMN}e^{i S_{\rm eff}[\gmn]}\nn\\
&=&-\frac{2}{\sqrt{-g}}\, \frac{\d S_{\rm eff}}{\d\gMN}\, .
\ees
The effective action defined in \eq{defSeff} is the relevant quantity for describing the physics at an energy scale that is sufficiently low compared to the mass of all matter fields, so that the matter fields can be integrated out. In FRW the relevant scale is given by the Hubble parameter $H(t)$ at the time $t$ of interest. Thus,  one is  implicitly assuming that all matter fields are very massive with respect to the Hubble scale, $m\gg H(t)$. However, if the theory includes a scalar field $\phi$ with  mass $m\,\lsim\, H(t)$, this assumption is violated since such a field cannot be integrated out completely. In particular, in the presence of an ultra-light scalar with 
$m\sim H_0$, the assumption is violated for all times $t\leq t_0$ (where $t_0$ denotes the present age of the Universe).
More precisely, at time $t$ we can integrate out the Fourier modes $\phi_{\bf k}$ (where ${\bf k}$ denotes the comoving momentum, related to the physical momentum 
by ${\bf k}_{\rm phys}={\bf k}/a(t)$)
whose frequency satisfies
\be\label{defhigh}
\omega_ k\equiv \sqrt{m^2+(k^2/a^2)}\gg  H(t)\, .
\ee
The low-frequency modes with ${\bf k}$ such that $\omega_{k}\,\lsim \, H(t)$ remain in the effective action and should be treated as a  classical background, on the same footing as the FRW metric $\gmn$.\footnote{Our approach here differs from that in \cite{Parker:1999td}, where the authors introduce an ultralight scalar field with $m\sim H_0$, and then derive an effective action for the gravitational field only, by integrating out all modes of the scalar field, including its low-frequency modes.} 
It is then  convenient to split the field $\phi(x)$ into its low- and high-frequency parts. We 
first expand $\phi(x)$ in Fourier modes,
\be\label{chi}
\phi(x)=
\int\frac{d^3k}{(2\pi)^3 \sqrt{2k}}\, 
\[
a_{\bf k}\phi_{\bf k}(t) e^{i{\bf k\cdot x}}+
a_{\bf k}^{\dagger}\phi^*_{\bf k}(t) e^{-i{\bf k\cdot x}}\]\, ,
\ee
and define
\be
\phi (x)=\chi(x)+\varphi(x)
\ee
where in $\chi (x)$ the integral over $d^3k$ runs only over the comoving momenta
with $|{\bf k}|<k_*(t)$ while in $\varphi (x)$ it runs only over comoving momenta with 
$|{\bf k}|>k_*(t)$. The value $k_*$ that separates the two regimes can be defined by rewriting \eq{defhigh} as
$\omega_{k_*}=\nu H(t)$ where $\nu$ is a dimensionless constant 
with $\nu\gsim 1$.\footnote{As usual in this separation the constant $\nu$ is in principle arbitrary, and the evolution of the low-energy effective theory with respect to $\nu$ is described by renormalization group equations.} Therefore
\be
k_*(t) = a(t)\sqrt{\nu^2 H^2(t)- m^2}\, .
\ee
Of course, the low-momentum modes exist only if $m$ and $t$ are such that $m\,\leq \, \nu H(t)$. At times where  $m\gg H(t)$, all modes become heavy and $\phi$ can be integrated out completely, or equivalently $k_*(t)=0$.
Observe that the separation between $\chi$ and $\varphi$ is time-dependent.  Denoting all other massive matter fields collectively by $\psi$, the effective action now becomes a functional of both the metric and $\chi$, given by

\begin{figure}[t]
\centering
\includegraphics[width=0.8\columnwidth]{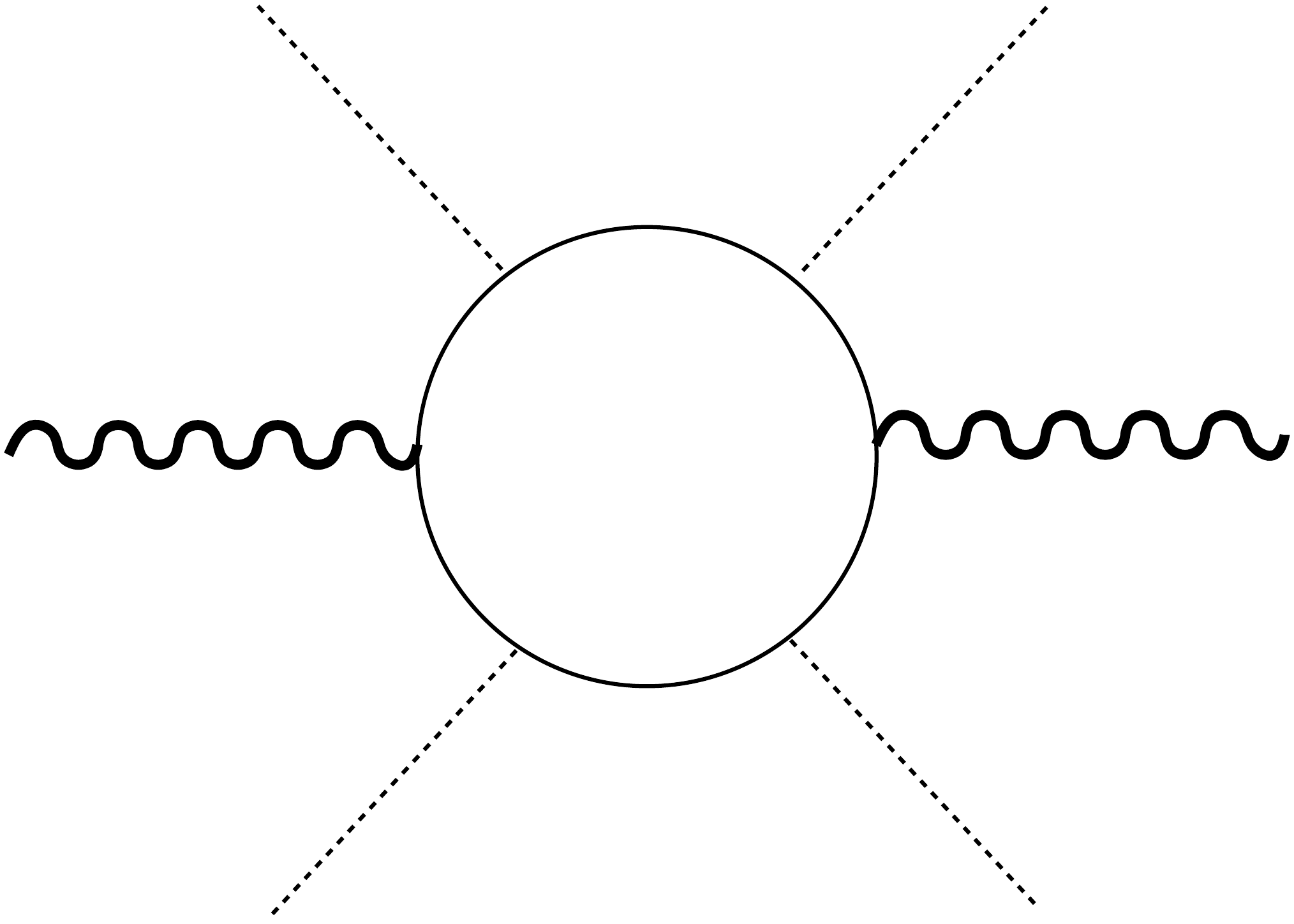}
\caption{\label{fig:loop}
A Feyman diagram that contributes to the generation of a term $f(\chi)R$ in the effective action. The  wavy line represents the classical gravitational field, the solid line in the loop
is the field $\varphi$ and the dashed line represents the field $\chi$. A graph with $n$ external $\chi$ lines contributes  a term $\propto\chi^nR$ to the effective action (here $n=4$).
}
\end{figure}

\be\label{Seffchi}
e^{i S_{\rm eff}[\gmn,\chi]}=\int {\cal D}\psi{\cal D}\varphi
\,
e^{iS_m[\gmn,\chi+ \varphi,\psi]}\, .
\ee
In principle
$S_{\rm eff}[\gmn,\chi]$ will include all possible terms that are consistent with general covariance.

In particular,  there can be a term proportional to 
$\int d^4x\sqrt{-g}\, f(\chi)R$, where $f(\chi)$ is some function whose explicit form depends on the details of the original action. Diagrammatically, this term comes from graphs such as that in fig.~\ref{fig:loop}, where the wavy lines represent the background gravitational field, the solid line circulating in the loop represents the high-frequency modes $\varphi$ and the external dashed lines represent $\chi$. A graph with the insertion of $n$ external $\chi$ lines contributes  a term proportional to $\chi^nR$ to the effective action. Observe that the $\chi$ field only occurs in external lines, since 
it is not integrated over.
Further notice that a graph with two external wavy lines produces the term in $R$ quadratic in the background metric. Terms cubic and of higher order in the background metric coming from $R$ are obtained from graphs with three or more external wavy lines. The integral in the  loop runs only over high-frequency momenta, i.e. $k_*<|k|<\infty$, and  can be regularized in the UV using e.g. dimensional regularization.

The graphs shown 
in  fig.~\ref{fig:loop} assumes the existence of a $\phi^3$ interaction term in the potential $V(\phi)$. After splitting $\phi=\chi+\varphi$ this produces a vertex
$\chi\varphi\varphi$. With a $\lambda\phi^4$ term in the potential we had vertices such as $\chi\chi\varphi\varphi$ instead, where two external $\chi$ lines are attached at each vertex in the loop. Furthermore, the gravitational field must be similarly split as $\gmn=\gBmn+\hmn$, where the background $\gBmn$ contains the low frequency modes, and basically coincide with the background FRW metric, while $\hmn$ contains the high-frequency modes and describes quantum fluctuations over the background. In \eq{Seffchi} we must therefore integrate also over $\hmn$, so
there are also graphs such as that in fig.~\ref{fig:loop} in which the internal $\varphi$ loop is replaced by a loop where $\hmn$ circulates. The non-linear terms in the gravitational action provide  vertices
such as $\bar{g}hh$  that allow us to attach  external $\gBmn$ lines to the $\hmn$ loop. Similarly  the interaction of $\phi$ with gravity provides 
vertices  such as ${\chi}hh$ that allows us to attach external $\chi$ lines to the loop. Therefore terms of the form $f(\chi)R$ are generated through the gravitational interaction even if
$V(\phi)=(1/2)m^2\phi^2$ and  there is no direct interaction between $\chi$ and $\varphi$.
All these graphs are generically non-zero, and therefore a term $f(\chi)R$ is in general generated in the effective action, even if  there was no term $f(\phi)R$ in the original action for $\phi$.

Among all terms that can be generated by the integration over the high-frequency modes, 
at the phenomenological level the term $f(\chi)R$  is the most interesting one, since it involves the lowest number of derivatives. Therefore it is  not suppressed at scales below the Planck mass, contrary to terms involving $R^2$, $\Rmn\RMN$, etc. or interaction terms involving derivatives of $\chi$. 
Therefore the most general effective action,  for generating a cosmologically relevant $\cav\Tmn\vac$,
is of the form
\bees\label{SchiJordan}
S_{\rm eff}[\gmn,\chi]&=&\int d^4x\, \sqrt{-g} \\
&&\times \[ \frac{1-f(\chi)}{16\pi G}R 
-\frac{1}{2}\gMN\pam\chi\pan\chi -V(\chi)\]\, ,\nn
\ees
plus higher-derivative operators, that are not relevant at scale much below the Planck scale, and possibly non-minimal kinetic terms, as in k-essence models.

The integration over the high-frequency modes in FRW  has also been  studied, in a somewhat different language,  in the context of stochastic inflation \cite{Starobinsky:1986fx,Morikawa:1989xz,Calzetta:1995ys}. For instance,  for a minimally coupled scalar field $\phi$  in De~Sitter space, the result for $S_{\rm eff}[\gmn,\chi]$
can be written in the form~\cite{Morikawa:1989xz}
\be\label{Seffxi}
e^{iS_{\rm eff}[\gmn,\chi]}=\int{\cal D}\xi \, P[\xi] e^{iS_0[\gmn,\chi,\xi]}
\ee
where $S_0$ is an action from which one derives the classical equation of motion for $\chi$, in the form
\be
\ddot{\chi}+3H\dot{\chi}-a^{-2}\n^2\chi+V'(\chi)=\dot{\xi}+3H\xi\, .
\ee
Here $\xi$ is a random field with a two-point correlator
\be
\langle\xi(x)\xi(x')\rangle =A(x-x')\, ,
\ee
and $P[\xi]$ in \eq{Seffxi} is given by
\be
P[\xi]=\exp\left\{-\frac{1}{2}\int d^4xd^4x'\, \xi(x)A^{-1}(x,x')\xi(x')\right\}\, .
\ee
Defining again the high-frequency modes as the modes with $\omega_k> \nu H$, the form of $A(x-x')$ in the most interesting limit $\nu\gsim 1$ has not been computed, while 
in the limit $\nu\ll 1$ it is given by~\cite{Morikawa:1989xz}
\be
A(x-x')=\frac{H^3}{4\pi^2}\,\frac{\sin(\nu H a r)}{\nu H a r}\, \d(t-t') +{\cal O}(\nu^2)\, ,
\ee
where $r=|\vx-\vx'|$. The above formulation is physically quite intuitive as it allows us to interpret $\chi$ as a classical field driven by some stochastic noise. This is the typical picture of stochastic inflation. If one was to integrate out $\xi$ from \eq{Seffxi} 
one would find the effective action
$S_{\rm eff}[\gmn,\chi]$ that can be directly obtained from diagrams such as that in
fig.~\ref{fig:loop}.

It is also instructive to directly compute
$\cav\Tmn\vac$ based on the sum of zero-point energies,
and to see how a term  $f(\chi)R$ could be generated in the effective action even if it was not present at first.
In the usual setting with only massive fields, e.g.\ a massive scalar field $\Phi$,  one expands the field in terms of creation and annihilation operators
\be\label{phi}
\Phi(x)=
\int\frac{d^3k}{(2\pi)^3 \sqrt{2k}}\, 
\[
a_{\bf k}\Phi_{k}(t) e^{i{\bf k\cdot x}}+
a_{\bf k}^{\dagger}\Phi^*_{k}(t) e^{-i{\bf k\cdot x}}\]\, .
\ee
The modes $\Phi_k(t)$ are obtained by solving the equation of motion in the given FRW metric, here
\be\label{eqvarphi}
\Phi_k''+2\frac{a'}{a}\Phi_k'+k^2\Phi_k=0\, ,
\ee
where the prime denotes derivatives with respect to conformal time $\eta$.
Then the vacuum energy density is given by
\be\label{rhophikd3k}
\cav T_{00}\vac =\frac{1}{2}\int\frac{d^3k}{(2\pi)^3 2k}\, 
\( |\dot{\Phi}_k|^2+\frac{k^2}{a^2}|\Phi_k|^2\)\, ,
\ee
where some regularization of the momentum integral is understood, e.g.\ a cutoff $\lc$ over momenta or dimensional regularization.
Thus, according to the standard rules of semiclassical gravity, the field $\Phi$ is treated as a quantum field, while the metric $\gmn$ is treated classically, and the modes of the quantum field 
are determined solving an equation in this classical background. The energy density
$\cav T_{00}\vac$ calculated in this way depends explicitly on the Hubble parameter $H(t)$ because the modes $\Phi_k(\eta)$ depend on the background metric, in this case on the scale factor $a(\eta)$ and conformal time $|\eta|\sim (a H)^{-1}$ of FRW. For example, in a De~Sitter epoch or during matter dominance (MD), 
\be\label{phiketa}
\Phi_k(\eta)=\frac{1}{a(\eta)}\(1-\frac{i}{k\eta}\) e^{-ik\eta}\, .
\ee
In the   limit $k\eta\ra\infty$, and therefore $k_{\rm phys}=k/a\gg H$, these modes reduce to the usual plane waves of flat space (this corresponds to the most natural choice of vacuum state, namely the Bunch-Davies vacuum; the most general form of the modes is a superposition of positive and negative frequency modes with arbitrary Bogoliubov coefficients). Therefore the leading UV divergence in \eq{rhophikd3k} is the same as in flat space, i.e.\ the $\lc^4$ term if we regularize with a cutoff in momentum space. However, the term
$i/k\eta$ in \eq{phiketa}, as well as the time dependence in the factor $1/a(\eta)$, affect the subleading divergence, and produce a quadratic divergence $\propto H^2(t)\lc^2$. (For details on these calculations see e.g.\ ref.\ \cite{Maggiore:2010wr}.)

When there is also an ultra-light scalar field  $\phi$ the above scenario changes, because $\phi$ must be split into its low-frequency modes $\chi$ and its
high-frequency modes $\varphi$. The field $\chi$  should be treated as classical on the same footing as the metric $\gmn$ since, on physical grounds, modes with wavelength larger than the horizon are expected to undergo decoherence. (See ref.\ \cite{Calzetta:1995ys} for a more careful discussion of this point.) The high-frequency modes $\varphi$ should instead be quantized in the classical background, given both by $\gmn$ and $\chi$. In particular, the equation of motion for $\varphi_k$ that replaces \eq{eqvarphi} will depend both on the scale factor and on the homogeneous field $\chi(\eta)$. It is derived from the equation of motion of $\phi$ by setting $\phi =\chi +\varphi$ and expanding $\varphi$ in momentum modes $\varphi_k$. As a result the modes (\ref{phiketa}) will acquire a dependence on $\chi$ which will generically translate into a $\chi$-dependence of 
$\cav\Tmn\vac$. In general, this produces all terms which are consistent with the symmetries of the problem, and therefore also terms corresponding to $f(\chi)R$ in the effective action.

Finally, we observe that similar considerations hold for massless conformally-coupled fields, including the electromagnetic field. If it were not for the conformal anomaly, in FRW (which is conformally equivalent to Minkowski space)  these fields would decouple from the metric and could be integrated out trivially. Because of the conformal anomaly  at the one-loop level these fields do not decouple but  still they can be integrated out exactly, giving rise to the anomaly-induced effective action. However the fact that, technically, it is possible to integrate them out exactly does not mean that, physically, this is the correct thing to do. In any case, the long-wavelength modes of these fields should be kept in the effective action, on the same footing as the classical metric, and produce a classical background (whether of a scalar field, of an electromagnetic field, etc.) which in general will  be relevant on cosmological scales. Excluding from the integration the modes with 
physical momenta $k_{\rm phys}<k_*/a(t)\sim H(t)$ could also in principle affect  the non-local terms of the anomaly-induced effective action.

\subsection{Interacting vacuum fluctuations}

We next discuss the properties of $\cav\Tmn\vac$ derived from the effective action
(\ref{SchiJordan}). Models of this form have been studied in the literature as generalizations of the simplest quintessence models \cite{Uzan:1999ch,Amendola:1999qq,Chiba:1999wt,Perrotta:1999am,Chiba:2001xx,Chiba:2010cy}. Of course such models can be introduced and investigated in their own right, without being related to zero-point fluctuations. Our line of arguments leading to such models,  based on the most general form of $\cav\Tmn\vac$ in the presence of ultra-light scalar fields, gives  further motivation for their study. This insight  will also allow us the recast the cosmological equations of the model in an interesting form, 
which can be seen to describe the interaction between vacuum fluctuation of the field  $\phi$ and the classical part $\chi$ of the same field.

Specializing to a FRW metric and to a spatially homogeneous field
$\chi=\chi(t)$, 
the equations of motion derived from the action (\ref{SchiJordan}) 
are~\cite{Uzan:1999ch,Amendola:1999qq,Chiba:1999wt,Perrotta:1999am,Chiba:2001xx,Chiba:2010cy}
\bees\label{FriedchiZ}
H^2&=&\frac{8\pi G}{3}(\rho_B+\rho_{\chi})\, ,\label{Friedbchi}\\
2\dot{H}+3H^2&=&-8\pi G (p_B+p_{\chi})
\, ,\label{eqii}\\
\ddot{\chi}+3H\dot{\chi}+V'(\chi)&=&- \frac{3}{8\pi G}(\dot{H}+2H^2)f'(\chi)
\, ,
\label{chiddot}
\ees
where $\rho_B,\,p_B$ are the energy density and pressure of the matter-radiation background, respectively, and $\rho_\chi,\,p_\chi$ are those associated to $\chi$ as given below. The prime denotes the derivative with respect to $\chi$. Denoting the total EOS parameter by $w_{\rm tot}\equiv(p_B+p_\chi)/(\rho_B+\rho_\chi)$, the two Friedmann equations are combined to find
\be\label{dotH32}
\dot{H}=-\frac{3}{2} H^2 (1+w_{\rm tot})\, ,
\ee
such that \eq{chiddot} can be recast as
\be\label{chiddot2}
\ddot{\chi}+3H\dot{\chi}+ V'(\chi)=-\frac{3}{16\pi G} H^2 (1-3w_{\rm tot})f'(\chi)
\, .
\ee
It is convenient to separate $\rho_{\chi}$ and $p_{\chi}$ into two parts,
\be\label{split}
\rho_{\chi}=\rho_X+\rho_Z\, ,\qquad
p_{\chi}=p_X+p_Z\, ,
\ee
where
\be\label{defrhoXpX}
\rho_X=\frac{1}{2}{\dot{\chi}}^2+V(\chi)\, ,\qquad
p_X=\frac{1}{2}\dot{\chi}^2-V(\chi)\, ,
\ee
are the standard contributions of the classical part, while
\bees
\hspace*{-3.8mm}\rho_Z&=&\frac{3}{8\pi G}  \[H^2f(\chi)  +H\dot{f}(\chi)\]\label{rhoZbchi} \, ,\\
\hspace*{-3.8mm}p_Z&=&-\frac{1}{8\pi G}\[
(2\dot{H}+3H^2)f (\chi)+2H\dot{f}(\chi)+\ddot{f}(\chi)\]\hspace*{-1.5mm}.
\label{pZbchi}
\ees
Using again \eq{dotH32},  \eq{pZbchi} can also be rewritten as
\be\label{pbchiZ2}
p_Z=\frac{1}{8\pi G}\[ 3w_{\rm tot}H^2f(\chi)-2H\dot{f}(\chi)-\ddot{f}(\chi) \]\, .
\ee
The subscript $Z$, which stands for ``zero-point fluctuations", stresses that in our approach 
$\rho_Z$ and $p_Z$ can be interpreted as a contribution coming from
vacuum fluctuations, i.e.\ from the VEV $\cav\Tmn\vac$ obtained through
\eq{TmnSeff} from an effective action that includes the contribution $f(\chi)R$ which generically appears in the presence of the ultra-light field $\chi$. 

A rather interesting property of $\rho_Z$ is that it is of the order of the critical density at all times. For instance, recalling that the critical density at time $t$ is 
$\rho_c(t)=(3/8\pi G) H^2(t)$, we see that
the term $(3/8\pi G)H^2(t)f(\chi)$ in 
$\rho_Z$ is just $f(\chi)\rho_c(t)$.  In  the slow-roll
regime $\chi(t)$  is roughly constant, and this term  provides an approximately fixed fraction of the critical energy density. In this way we obtain a term of order  $\mpl^2H^2(t)$ in the energy density, which is not reabsorbed into a renormalization of Newton's constant because its prefactor contains the function $f(\chi)$.

The above equations can be rewritten in a form which allows for  an appealing physical interpretation, namely that the ``classical" energy-momentum tensor described by $\rho_X$ and $p_X$ interacts with the vacuum fluctuations, described by $\rho_Z$ and $p_Z$. The interaction is governed by a coupled energy-momentum conservation equation (thereby providing an explicit realization of a mechanism proposed 
in \cite{Maggiore:2010wr,Maggiore:2011hw}).
To this purpose, it is convenient to transform the time derivatives into derivatives with respect to $x\equiv \ln a$, which we denote by $\px$. By means of \eq{dotH32} we can then write
\bees
\label{rhoff}
\rho_Z&=&\rho_c\( f + \px f\) \, ,
\\
p_Z&=&w_Z\rho_Z\, ,
\\
\label{wZ}
w_Z&=&w_{\rm tot} -\frac{(1+3w_{\rm tot})\px f + 2\px^2 f}{6(f+\px f)}\, .
\ees
If the field $\chi$ does not exchange energy with the other forms of matter included in the background action $S_B$ (as we have indeed already assumed when deriving \eq{chiddot}), its energy density and pressure satisfy the conservation equation
\be
\dot{\rho}_{\chi}+3H(\rho_{\chi}+p_{\chi})=0\, .
\ee
Splitting $\rho_{\chi}$ and $p_{\chi}$ as in \eq{split} and defining
\be\label{wbchi0}
w_X\equiv \frac{p_X}{\rho_X}=
\frac{(1/2){\dot{\chi}}^2-V(\chi)}{(1/2){\dot{\chi}}^2+V(\chi)}\, ,
\ee
we can write the continuity equation as
\bees
\dot{\rho}_Z+3H (1+w_Z)\rho_Z&=&Q\, ,\label{Q1}\\
\dot{\rho}_X+3H (1+w_X)\rho_X&=&-Q\, ,\label{Q2}
\ees
where $Q(t)$ is the energy density transfer rate from $X$ to $Z$. Inserting the explicit expressions for $\rho_Z$ and $w_Z$ given in \eqs{rhoff}{wZ} we find
\be\label{QHrho}
Q=\frac{1}{2}(1-3w_{\rm tot})H(t)\rho_c(t)\,\px f\, .
\ee
Since $\rho_c(t)$ and $H(t)$ provide the natural scales for the energy density and for the
inverse of time, respectively, a natural scale for the rate of energy transfer is $H(t)\rho_c(t)$. We see from \eq{QHrho} that, with respect to this scale, the actual rate $Q$ differs by a factor proportional to $\px f$ and to $(1-3w_{\rm tot})$. Since $H\px f=\dot{f}$, we see that the actual rate of energy transfer is governed by the time variation of $f(x)$, and therefore in a slow-roll phase
$|Q|\ll H(t)\rho_c(t)$. Furthermore,
during RD $w_{\rm tot}\approx 1/3$, providing a further suppression factor.
However, note that $Q$ does in general not vanish exactly in RD because of the contributions from $p_Z$ and matter. We will see this for a typical example in the next section.

Conservation equations with energy exchange such as \eqs{Q1}{Q2}, that even go back to Bronstein (1933) 
(see \cite{Peebles:2002gy}), are at the basis of much recent work on interacting dark energy/matter models (see 
e.g.\ \cite{Wetterich:1994bg,Amendola:1999qq,CalderaCabral:2009ja,Valiviita:2009nu,Basilakos:2009wi,Grande:2011xf,Basilakos:2011wm} and references therein). Typically the functional form of $Q(t)$ is chosen on purely phenomenological grounds. It is interesting to see that in this model the form of $Q(t)$ is explicitly predicted in terms of $\px f$.\footnote{To make the connection with the discussion just after \eq{wwtot} it is useful to 
introduce the EOS parameter $w_*(t)$ defined by the condition
$\dot{\rho}_Z+3H(1+w_*)\rho_Z\equiv 0$. Using
\eqs{rhoff}{dotH32} we  get
$w_*=w_{\rm tot}-(\px f+\px^2f)/[3(f+\px f)]$. Observe that $\px f=0$ implies $w_*=w_{\rm tot}$ consistent with \eq{wwtot}. Comparing with \eq{wZ} we see that 
$w_Z=w_*+\bar{w}$ where $\bar{w}= [1-3w_{\rm tot}] \px f / [6(f+\px f)]$ and therefore in general $w_Z\neq w_*$.}

In summary, the energy-momentum tensor of    the classical part,
$T_{\mu\nu}^0=(\rho_X,\, a^2\d_{ij}p_X)$ and that of the vacuum fluctuations
$\cav\Tmn\vac=(\rho_Z,\, a^2\d_{ij}p_Z)$ are
in general not separately conserved:
\be
\n^{\mu} \cav\Tmn\vac =-\n^{\mu} T_{\mu\nu}^0\neq 0\, .
\ee
Rather, they interact through a coupled energy-momentum conservation equation.

\subsection{Cosmological evolution and tracking dark energy}  \label{sec:cosmevol}

The cosmological evolution obtained from \eqst{FriedchiZ}{chiddot}
has been studied in detail, see e.g.\ refs.~\cite{Uzan:1999ch,Amendola:1999qq,Chiba:1999wt,Perrotta:1999am,Chiba:2001xx,Chiba:2010cy}. To illustrate the typical evolution, we consider a model with 
\be\label{defFV}
\frac{f(\chi)}{16\pi G}\equiv F(\chi)=\frac{1}{2}\xi\chi^2\, ,\qquad
V(\chi)=\frac{1}{2}m^2\chi^2\, , 
\ee
and we take for definiteness $\xi>0$.
It is useful to introduce  the dimensionless parameter
\be
\mu \equiv \frac{m}{\xi^{1/2}H_0}\, . 
\ee
The ratio of the term $6H^2F$ in $\rho_Z$ to the potential $V$ is equal to $6H^2/(\mu^2H_0^2)$. Therefore, 
if $\mu=3$, these two terms become comparable at the present epoch, while in the past
$6H^2F\gg V$. For larger values of $\mu$ the potential $V$ begins to give a  significant contribution to the energy density somewhat earlier than today, while for smaller $\mu$ it will only become important in the future. The evolution of $\chi$, determined by \eq{chiddot2}, is different depending on whether 
$V'(\chi)$ or $3H^2 (1-3w_{\rm tot})F'(\chi)$ dominates. The ratio of these two terms is
\be
\frac{3H^2 (1-3w_{\rm tot})F'(\chi)}{V'(\chi)}=\frac{3H^2 (1-3w_{\rm tot})}{\mu^2H_0^2}\, .
\ee
Since observationally $w_{\rm tot}^0\approx -0.7$ today, this ratio becomes of order unity at the present epoch for $\mu\simeq 3$. In the past it was much larger than one. 
This is true even in RD despite the fact that, during RD, $(1-3w_{\rm tot})\ll 1$. To see this in a simpler setting, lets us at first include just the contribution of matter to $w_{\rm tot}$ in the RD phase.
Including both matter and radiation, deep in  RD where $\rho_M\ll\rho_R$ and the scale factor $a\ll a_{\rm eq}$ (where $a_{\rm eq}\equiv \Omega_R^0/\Omega_M^0\approx3\times 10^{-4}$) we have
\be\label{wtotnof}
w_{\rm tot}=\frac{(1/3)\rho_R}{\rho_R+\rho_M}
=\frac{1}{3}\, \frac{1}{1+ a /a_{\rm eq}}\, .
\ee
Therefore
\be
(1-3 w_{\rm tot})  H^2\simeq \frac{a}{a_{\rm eq}} H^2 \, .
\ee
In RD $a\propto t^{1/2}$ and $H^2\propto 1/t^2$,  so
$(1-3 w_{\rm tot})  H^2\propto t^{-3/2}$ and, going backward in time, the growth of $H^2$ more than compensate the decrease of $ (1-3 w_{\rm tot})$. 

More importantly,  in this model  there is a significant contribution from DE to 
$w_{\rm tot}$ even in RD. This contribution, contrary to that of matter,  is not parametrically suppressed in the limit $a/a_{\rm eq}\ll 1$. To compute it in  the matter and radiation era we can safely neglect $p_X$. Then we find from the expressions in the last section
\be
1-3 w_{\rm tot} \simeq \frac{(1-f-\px f)/(1+a_{\rm eq}/a) +\px f +\px^2f}{1-f-\px f/2 } \,,
\ee
which correctly reduces to \eq{wtotnof}  when $f={\rm const.}$
Thus, the combination $(1-3w_{\rm tot})$ is generically non-zero, and
for $\mu={\cal O}(1)$ the evolution of $\chi$ during RD and most of MD is dominated by the non-minimal coupling $F(\chi)R$. The potential $V(\chi)$ only becomes relevant when approaching the current epoch, while for larger values of $\mu$ the potential becomes relevant earlier. Eventually, the field $\chi$ will exit the slow-roll phase and perform damped oscillations around the minimum of $V(\chi)$ at $\chi=0$.

\begin{figure*}[t]
\centering
\includegraphics[width=0.45\textwidth]{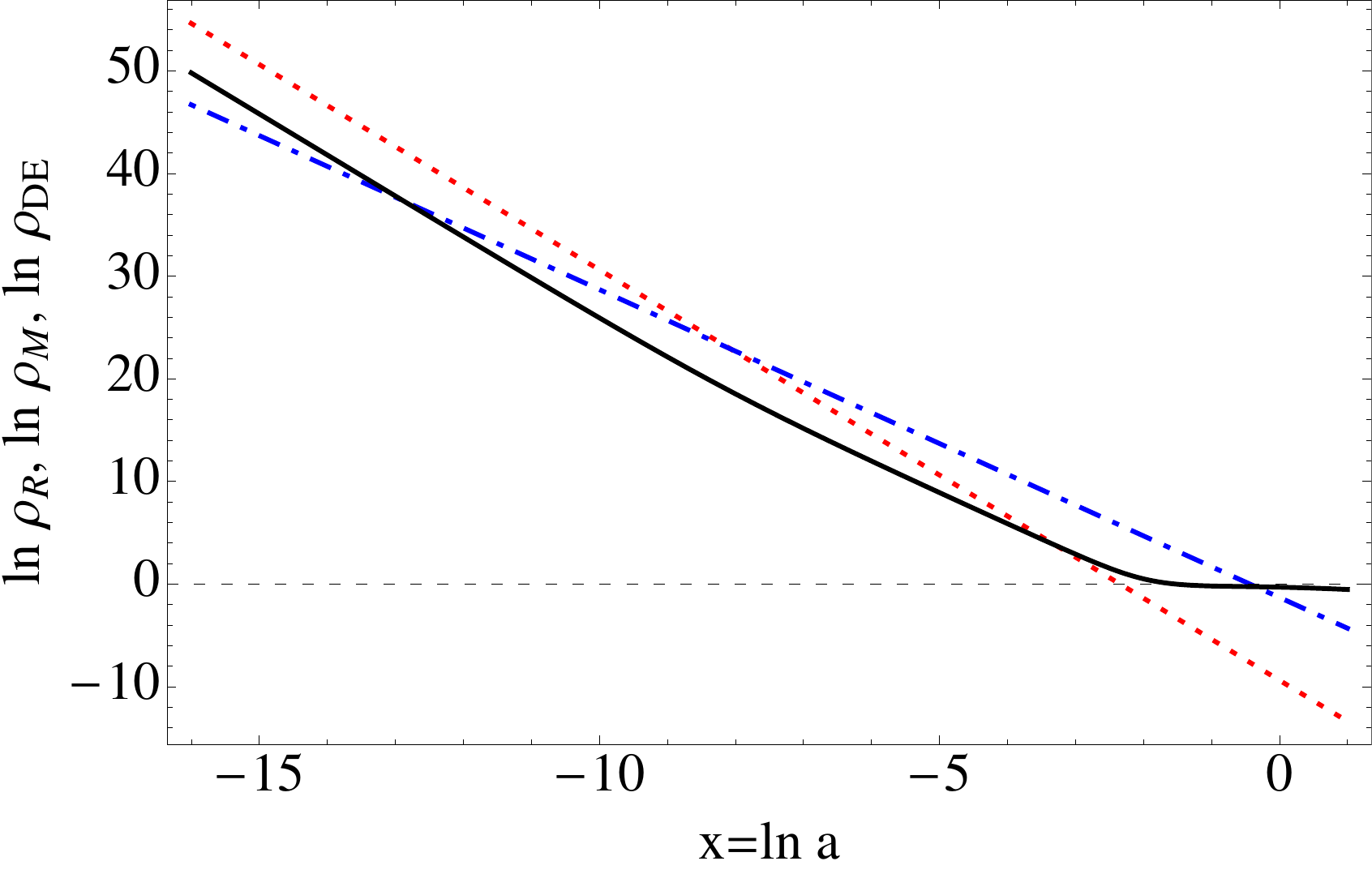}\quad
\includegraphics[width=0.45\textwidth]{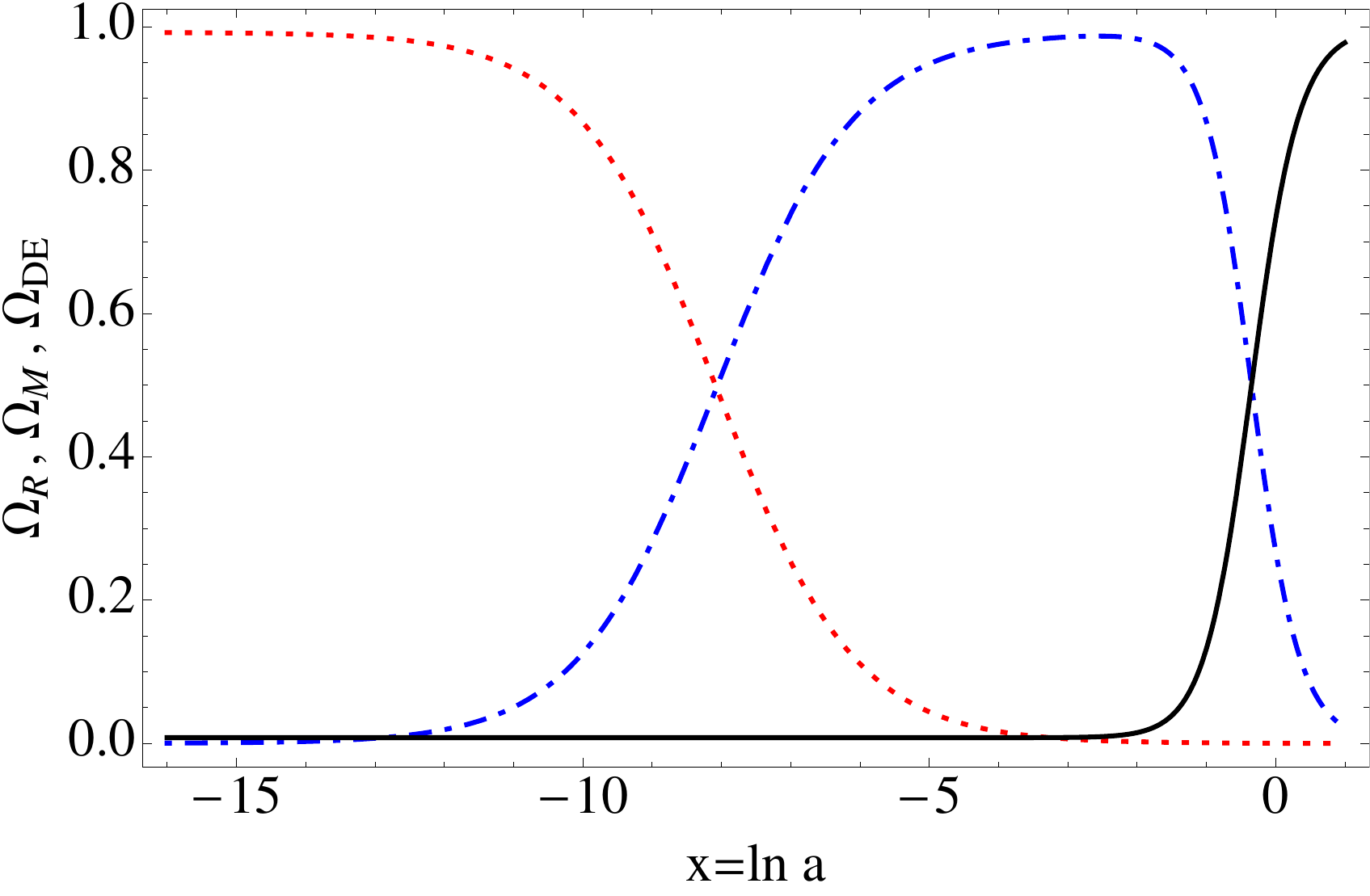}
\caption{\label{fig:logrho}
\emph{Left:} $\log\rho$ for radiation (red, dotted), matter (blue, dot-dashed) and dark energy (black solid) as functions of $\log a$.
\emph{Right:} the same plot on a linear vertical scale, in terms of $\Omega_i=\rho_i(t)/\rho_c(t)$, with $i=$ R, M, DE.
We used the values $\mu =24.3$, $\xi=5\times 10^{-4}$ and the initial conditions 
$\chi_{\rm in}=4$ (in units $8\pi G=1$), $\chi'_{\rm in}=0$ deep in RD.}
\end{figure*}

\begin{figure*}[t]
\centering
\includegraphics[width=0.45\textwidth]{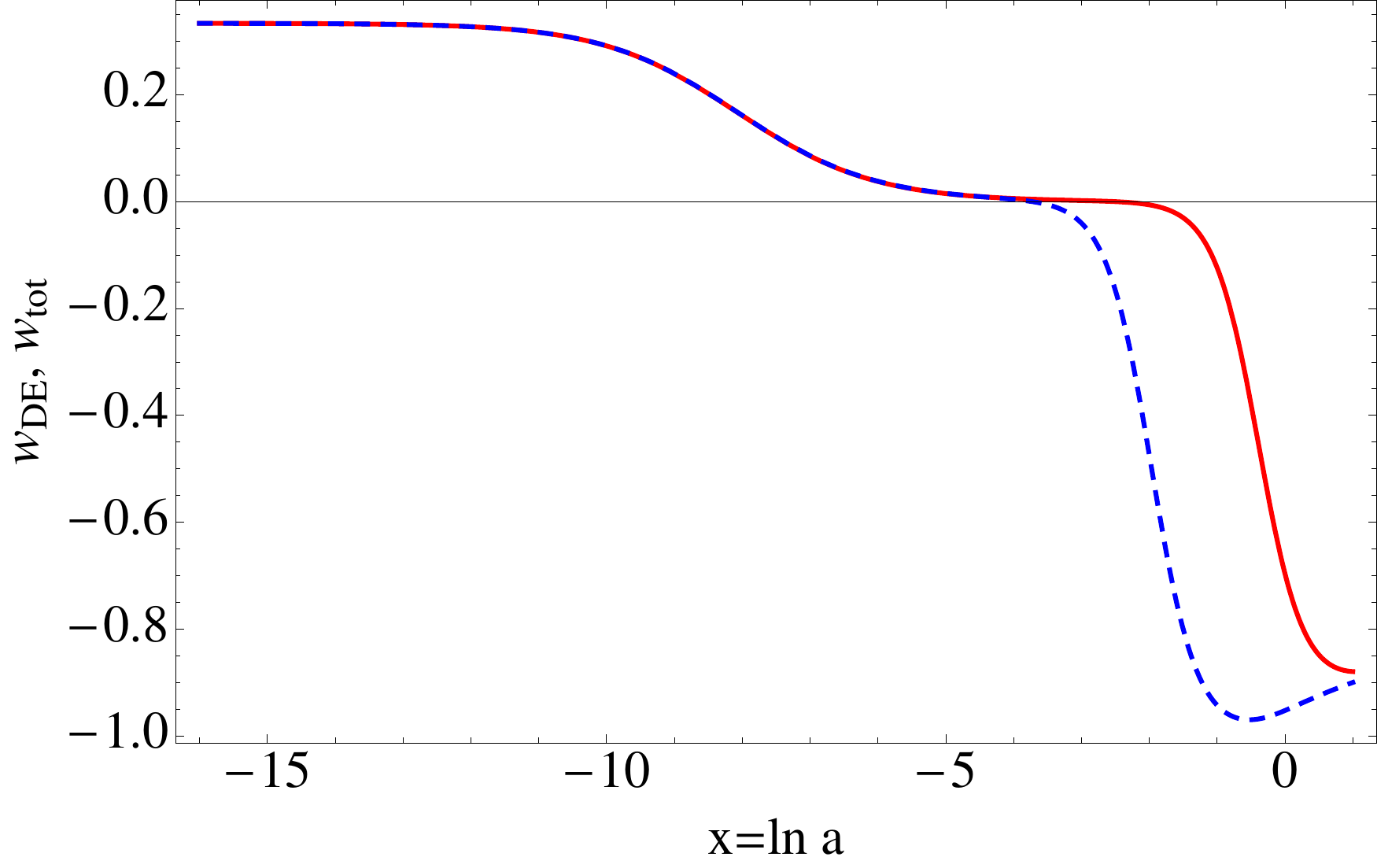}\quad
\includegraphics[width=0.45\textwidth]{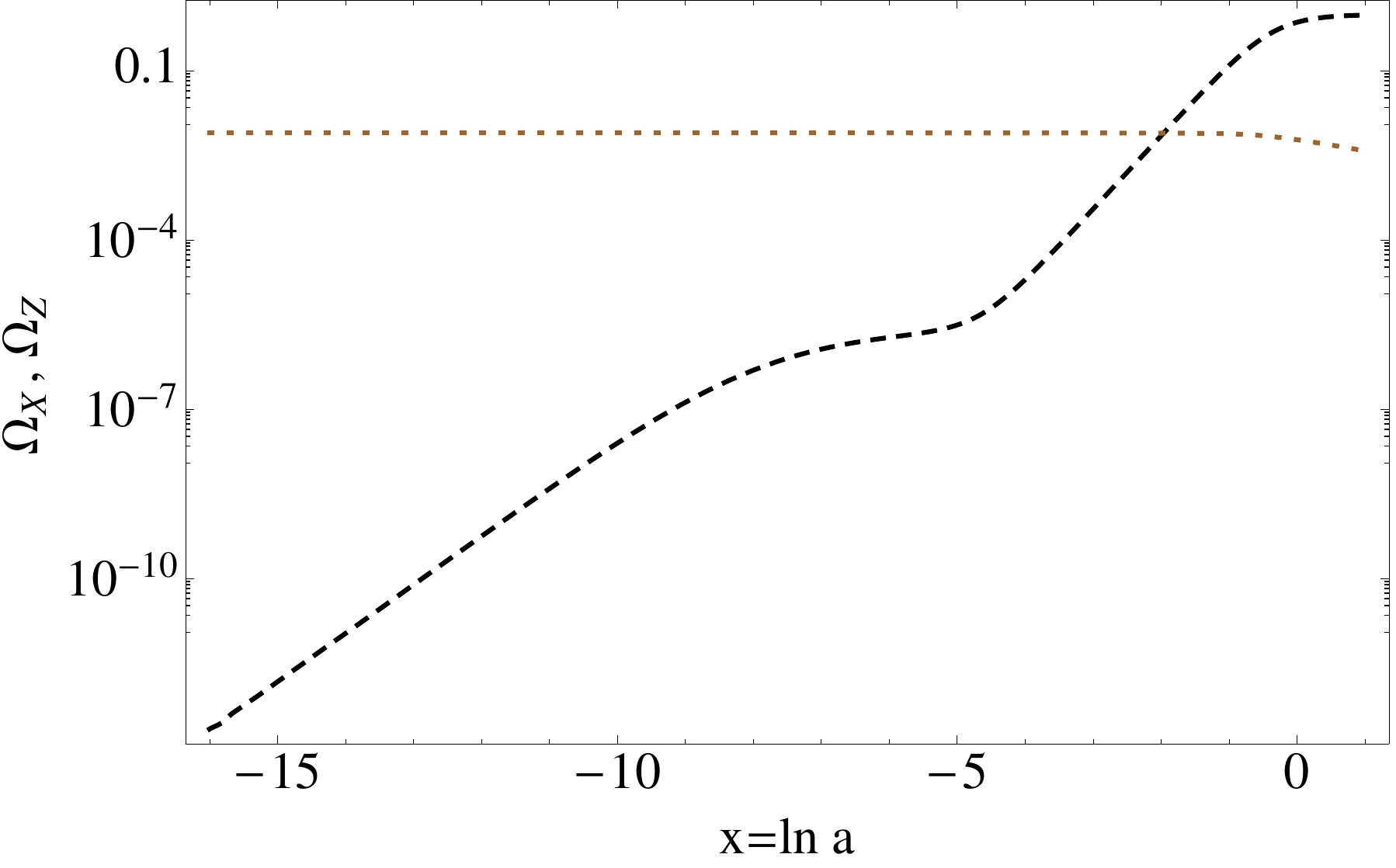}
\caption{\label{fig:ws}
\emph{Left:} $w_{\rm tot}$ (red, solid) and $w_{\rm DE}$ (blue, dashed) as functions of $\log a$ for the same parameters as in fig.~\ref{fig:logrho}.
\emph{Right:} the separate contributions $\Omega_X=\rho_X(t)/\rho_{\rm tot}(t)$ (black, dashed) and $\Omega_Z=\rho_Z(t)/\rho_{\rm tot}(t)$ (brown, dotted).}
\end{figure*}

In this model the total DE density $\rho_X+\rho_Z$ tracks the total energy density $\rho_c(t)$, as we can see from  fig.~\ref{fig:logrho}. 
This is basically due to the combination of two factors: during RD and deep into MD  the tracking is ensured by the term $f(\chi)\rho_c(t)$ in $\rho_Z$. This term, however, cannot be the dominant one today, since it is  proportional to $H^2(t)$, and in the recent cosmological epoch a time dependence $\rho_{\rm DE}\propto H^2(t)$ is observationally excluded. Thus, approaching the recent epoch, we must switch to a regime where $\rho_X$ dominates and $\chi$ is still approximately slowly-rolling such that  $w_X$ is close to $-1$, see \eq{wbchi0}, and the DE density is approximately constant.

With suitable values of  the parameters $\xi$, $\mu$ and of the initial conditions we can get values for $\Omega_{\rm DE}$ and $w_{\rm DE}$ today of the order of the observationally preferred ones. For instance, in fig.~\ref{fig:logrho} we used $\mu =24.3$, $\xi=5\times 10^{-4}$ and we set the initial conditions $\chi_{\rm in}=4$ and  $\chi'_{\rm in}=0$ deep in RD (measuring $\chi$ in units $8\pi G=1$). With these values we get $\Omega_{\rm DE}\simeq 0.73$ and $w_{\rm DE}=-0.952$ today. The evolution of the EOS parameter $w_{\rm DE}\equiv p_{\rm DE}/\rho_{\rm DE}$ is shown in the left panel of fig.~\ref{fig:ws}, together with the evolution of $w_{\rm tot}$. 
This shows that a realistic cosmological evolution can be obtained in this 
model.\footnote{Furthermore, in this model the limits on the time variation of Newton's constant require
$32\pi G(\xi\chi_0)^2<\omega_{\rm BD}^{-1}$, where $\chi_0$ is the present value of $\chi(t)$ and $\omega_{\rm BD}$ is the Brans-Dicke parameter~\cite{Perrotta:1999am}. Using the bound $\omega_{\rm BD}>4.2\times 10^4$ inferred by the Cassini mission
\cite{Will:2005va} and measuring $\chi$ in units $8\pi G=1$, this bound reads $\xi\chi_0< 2.4\times 10^{-3}$. This  is satisfied by the solution shown in fig.~\ref{fig:logrho}, for which
$\xi\chi_0\simeq 1.9\times 10^{-3}$.}

\begin{figure}[t]
\centering
\includegraphics[width=0.9\columnwidth]{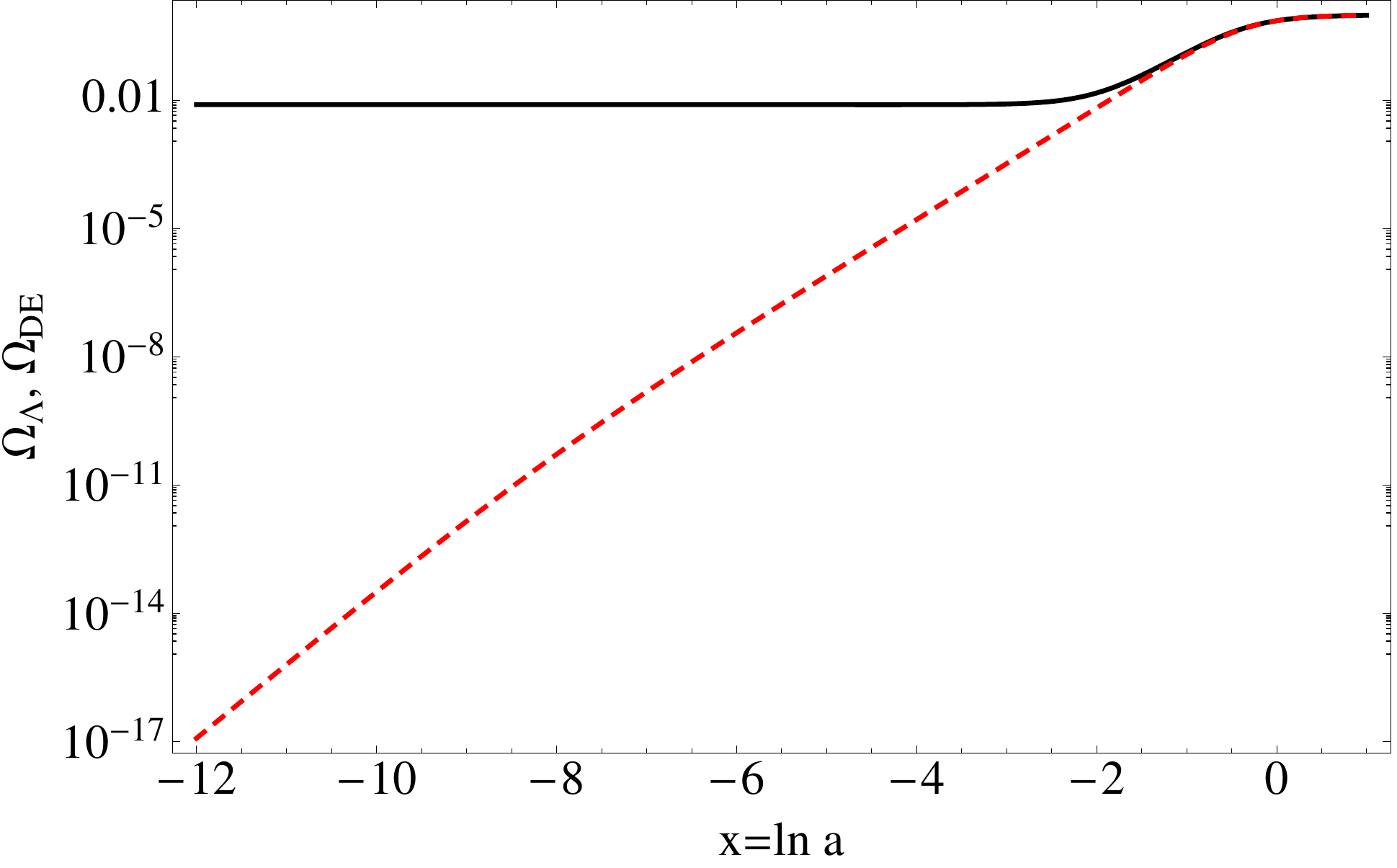}
\caption{\label{fig:rhoDElambda2}
The ratio $\Omega_{\rm DE}(t)\equiv \rho_{\rm DE}(t)/\rho_{\rm tot}(t)$, computed in the model with 
$F(\chi)$ and $V(\chi)$ given in \eq{defFV} (solid, black) compared to the  ratio 
 $\Omega_{\Lambda}(t)\equiv \rho_{\Lambda}/\rho_{\rm tot}(t)$
in $\Lambda$CDM (dashed, red).}
\end{figure}

It should however be stressed that the coincidence problem is not really solved by the fact that the DE density tracks the total energy. The total dark energy density $\rho_{\rm DE}=\rho_X+\rho_Z$ switches from an early regime where it is dominated by $\rho_Z$, and hence evolves approximately as $H^2(t)$, to a late regime where it is dominated by $\rho_X$, and therefore is approximately constant. The transition between these regimes can nicely be observed when looking at the separate evolution of $\rho_X(t)/\rho_c(t)$ and $\rho_Z(t)/\rho_c(t)$, plotted in the right panel of fig.~\ref{fig:ws} for the same model parameters used for the other figures. The time when the transition takes place is controlled by ratio between the mass of $\chi$ and the Hubble constant. In fact, it is the parameter $\mu=m/(\xi^{1/2}H_0)$ that is crucial here: the larger $\mu$ the earlier the transition. As we argued above, if $\mu=3$ the transition happens about at the current epoch. Notice though, compared to $\Lambda$CDM, the coincidence problem is still sensibly alleviated. This is clearly seen from 
fig.~\ref{fig:rhoDElambda2}, where the ratio $\rho_{\rm DE}(t)/\rho_{\rm tot}(t)$, computed in the model with 
$F(\chi)$ and $V(\chi)$ given in \eq{defFV}, is compared to the same ratio in $\Lambda$CDM. 

Exploring the parameter space to a certain extent, we find that the figures shown represent the typical behavior of this model, as long reasonable initial conditions are set. In principle, a deeper study of the stability and the parameter space of the system would be needed, but it is not the aim of this work to do so. The reason is the lack of motivation for the specific choices for $F(\chi)$ and $V(\chi)$ at this point. Therefore we will discuss a more general if slightly approximate approach in the following.

\subsection{Comparison with observations}

The model that we have discussed has a DE density $\rho_{\rm DE}(t)=\rho_X(t)+\rho_Z(t)$, where $\rho_X(t)$  is the typical quintessence energy density due to the rolling of the scalar field $\chi$ in the potential $V(\chi)$, while $\rho_Z(t)$ comes from the term $F(\chi)R$ in the effective action. With a suitable choice of the mass $m$ in $V(\chi)$ the term $\rho_X$ comes to dominate near the present epoch, while it is totally negligible at early times, and its EOS parameter $w_X$ is close to $-1$ today, as required by the observations. In contrast, $\rho_Z(t)$ is  proportional to $H^2(t)$. In order to have a viable model, $\rho_Z$ must be subleading today since a DE term proportional to $H^2(t)$ is observationally excluded (see e.g.
 \cite{Basilakos:2009wi,Grande:2011xf,Basilakos:2011wm}). However, the fact that this term is approximately a constant fraction of  $\rho_c(t)$ means that its integrated effect over the whole history of the Universe could give observable effects.
 
\subsubsection{A more general parametrization of $\rho_{\rm DE}$}\label{sect:more}
 
The aim of this section is to perform a comparison of such a model with cosmological observations, expanding on the results that we have presented in \cite{Maggiore:2011hw}. 
A possible way to proceed could be to choose a particular form of the functions $F(\chi)$ and $V(\chi)$, such as those given in \eq{defFV}, and find the corresponding limits on the parameters $\xi$ and $m$. However, such limits  would be tied to the specific choice made for the functions $F(\chi)$ and $V(\chi)$. In the absence of strong theoretical motivations for any such  choice, it seems more useful to extract some general features of the DE sector of such models, and parametrize the DE density in a form that might have a more general validity. This can be done by recasting \eqs{defrhoXpX}{rhoZbchi} in the form
\be\label{rhoVK}
\rho_{\rm DE}=V + H^2 \[\frac{1}{2}\(\px \chi\)^2 + 6(F + \px F) \]\, ,
\ee
where we have again transformed the derivative with respect to $t$ into the derivative with respect to $x\equiv\ln a$, denoted by $\px$. By means of the Friedmann equation, $H^2=(8\pi G/3)(\rho_R+\rho_M+\rho_{\rm DE})$, and the definition
\be
Z \equiv \frac{8\pi G}{3}\[\frac{1}{2}\(\px \chi\)^2 + 6(F + \px F) \]
\ee
we can write the DE density in the form
\be
\rho_{\rm DE} = \frac{V +(\rho_R +\rho_M) Z}{1-Z} \, .
\ee
Finally, we implicitly define the three functions $w_0(a)$, $\bar{\eps}_R(a)$ and $\bar{\eps}_M(a)$ and the constant $C$ by writing
\be\label{rhoepsRM}
\rho_{\rm DE} =  C a^{-3(1+w_0(a))} +\rho_R(a)\bar\eps_R(a) + \rho_M(a)\bar\eps_M(a)\, ,
\ee
such that
\bees\label{defeps}
\bar\eps_R(a)&=&\bar\eps_M(a)=\eps(a) \equiv \frac{Z}{1-Z} \,,\\
w_0(a) &=& -1 -\frac{1}{3}\ln\[\frac{V}{C Z}\epsilon(a)\] \,.
\ees
When $w_0$ is constant, the term $Ca^{-3(1+w_0)}$ in \eq{rhoepsRM} corresponds to the usual dark energy density of a minimal quintessence model. On the other hand, the presence of the terms $\rho_R\bar\eps_R$ and $\rho_M\bar\eps_M$ clearly displays the tracking behavior of the DE density. Choosing a specific model for the functions $F(\chi)$ and $V(\chi)$ corresponds to making a choice for $w_0(a)$ and $\eps(a)$. The constant $C$ is fixed by the DE density today, or better, by a combination of the model parameters in $F$ and $V$ and the initial value of $\chi$.

\begin{figure*}[t]
\centering
\includegraphics[width=0.45\textwidth]{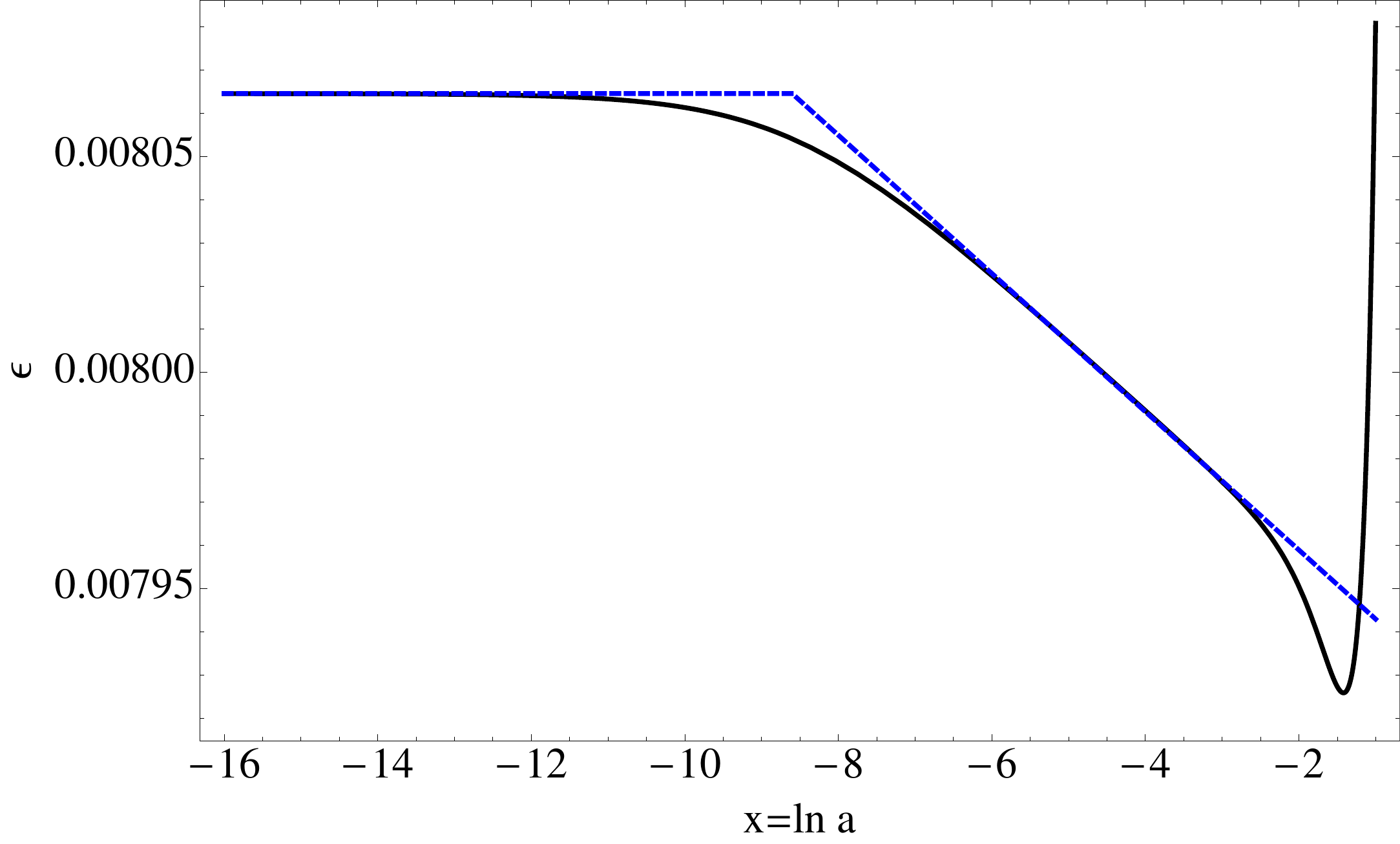}\quad
\includegraphics[width=0.45\textwidth]{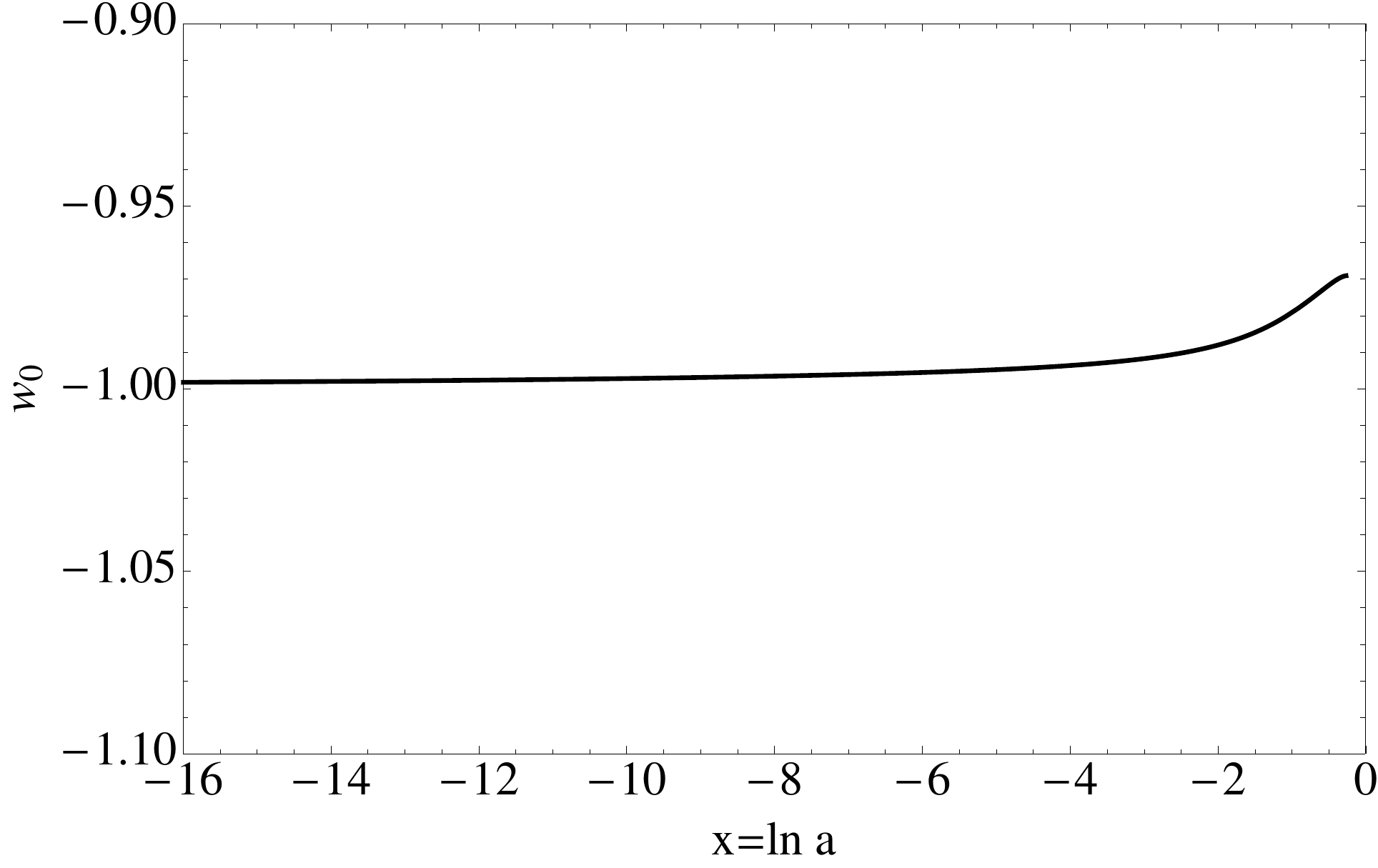}
\caption{\label{fig:eps2}
The evolution of the functions $\eps=Z/[1-Z]$ and $w_0$ defined in \eq{rhoepsRM} are shown for the model and parameters of Sec.~\ref{sec:cosmevol}, see fig.~\ref{fig:logrho}. In the left panel we also show the $\epsilon_R-\epsilon_M$ approximation with $\epsilon_R=0.00806$ and $\epsilon_M=0.00793$ (blue, dashed).}
\end{figure*}

Going back to the example discussed in the last section, $F(\chi)=(1/2)\xi\chi^2$ and $V(\chi)=(1/2)m^2\chi^2$, we solve \eqs{FriedchiZ}{chiddot} numerically and show the evolution of the functions $w_0(a)$ and $\eps(a)$ in fig.~\ref{fig:eps2}. We see that in the recent epoch where the term $Ca^{-3(1+w_0)}$ is relevant $w_0$ can roughly be approximated by a constant, within a few percent.  Moreover, the function $\eps(a)$ is constant to excellent approximation in RD and decreases linearly with $\ln a$ in MD.  This is the reason why we separated the terms into $\rho_R\bar\eps_R$ and $\rho_M\bar\eps_M$, as either one will be negligible when the other one is dominating. Although $\eps(a)$ raises again when the present epoch is approached, see fig.~\ref{fig:eps2}, at this stage the contribution from $\rho_M\bar\eps_M$ in the DE density is subdominant compared to the term 
$C a^{-3(1+w_0)}$ coming from $\rho_X$ (see the right panel in fig.~\ref{fig:ws}). Accordingly, the behavior of $\eps(x)$ in this regime is not very relevant. Therefore, we can approximately write
\bees
\bar\eps_R(a) &\simeq& \eps_R
\\
\bar\eps_M(a) &\simeq& \eps_M + B\ln a
\ees
with independent constant parameters $\eps_R$ and $\eps_M$. The constant $B$, on the other hand, is fixed by requiring the two branches to coincide at equality: $B=(\eps_R-\eps_M)/\ln a_{\rm eq}$.

The fact that $\bar\eps_R$ can be taken as constant is a consequence of $\chi$ rolling very slowly during RD, in fact it does essentially not move at all, see also ref.~\cite{Chiba:2010cy}. Thus, the function $Z(\chi)$ is virtually constant. The fact that in the MD era $\eps$ evolves linearly with $\ln a$ is because $\chi$ begins to roll, still slowly though, and therefore with $\ln a$ as this is the natural variable for cosmological evolution. Thus, this behavior appears to be quite general for typical viable models, and is likely to be not restricted to the specific model
$F(\chi)=(1/2)\xi\chi^2$ and $V(\chi)=(1/2)m^2\chi^2$ that we have studied here. We will therefore use
\eq{rhoepsRM} as our basic expression for the DE density, with $w_0(a)=w_0$ and
$\eps(a)$ given in terms of the constans $\eps_R$ and $\eps_M$.
This model then is a generalization of the one studied in
ref.~\cite{Maggiore:2011hw}, which corresponds to the limit $\eps_R=\eps_M$. It is a two-parameter generalization of $w$CDM, which in turn corresponds to the limit $\eps_R=\eps_M=0$. We will generically use the label $w$ZCDM for this generalization (including the limiting case $\eps_R=\eps_M$ studied in \cite{Maggiore:2011hw}).

\subsubsection{Results}

We estimate the model parameters with a full likelihood analysis using modified versions of the publicly available Markov-Chain Monte-Carlo sampling code CosmoMC \cite{Lewis:2002ah,Cosmomc} and the CMB-Boltzmann code CAMB \cite{Lewis:1999bs,Camb}. We assume a flat FRW background and put a prior on the age of the Universe to be between 10 and 20 Gyrs. Then we use the following datasets: constraints on the current Hubble parameter from Hubble Space Telescope (HST) observations \cite{Riess:2009pu} and on the number of relativistic degrees of freedom at Big-Bang Nucleosynthesis (BBN) \cite{Pisanti:2007hk}; the Union2 Compilation of 
Type Ia supernovae (SNe) 
of the Supernova Cosmology Project \cite{Amanullah:2010vv}; the angular power spectra of temperature and polarization anisotropies in the cosmic microwave background (CMB) from WMAP7 \cite{Komatsu:2010fb}, ACBAR \cite{Reichardt:2008ay} and CBI \cite{Sievers:2005gj}; and the Baryon Acoustic Oscillations (BAO) data from the Sloan Digital Sky Survey Data Release 7 (SDSS DR7) \cite{Percival:2009xn}.

\begin{table*}
\begin{ruledtabular}
\begin{tabular}{l l c c l c c l c c l c c}
\toprule
& & \multicolumn{2}{c}{$\Lambda$CDM} & & \multicolumn{2}{c}{$w$CDM }
  & & \multicolumn{2}{c}{$w$ZCDM1} & & \multicolumn{2}{c}{$w$ZCDM2}
\\
& & $\langle...\rangle$ & $\sigma$  & &  $\langle...\rangle$ & $\sigma$
  & & $\langle...\rangle$& $\sigma$  & &  $\langle...\rangle$& $\sigma$
\\  \\ \hline \\
\midrule 
$n_s$  & &  0.966 & 0.011  & &  0.963 & 0.013  & &  0.967 & 0.014  & &  0.971 & 0.016
\\ 
$\log[10^{10} A_s]$  & &  3.084 & 0.032  & &  3.088 & 0.032  & &  3.087 & 0.032  & &  3.084 & 0.034
\\
$\tau$  & &  0.087 & 0.014  & &  0.086 & 0.014  & &  0.087 & 0.015  & &  0.089 & 0.015
\\
$\Omega_b h^2 $  & &  0.023 & 0.001  & &  0.023 & 0.001  & &  0.0226 & 0.001  & &  0.0225 & 0.001
\\
$\Omega_{\rm DM} h^2 $  & &  0.113 & 0.003  & &  0.115 & 0.004  & &  0.115 & 0.005  & &  0.111 & 0.007
\\
$\theta$  & &  1.040 & 0.002  & &  1.040 & 0.002  & &  1.039 & 0.003  & &  1.042 & 0.006
\\
\midrule \vspace{1.5mm}
$w_0$  & &  - & -  & &  -1.043 & 0.079  & &  -1.067 & 0.087  & &  -1.11 & 0.11
\\
$\epsilon$  & &  - & -  & &  - & -  & &  0.0106 & 0.017  & &  - & -
\\
$\epsilon_M$  & &  - & -  & &  - & -  & &  - & -  & &  0.041 &  0.048
\\
$\epsilon_R$  & &  - & -  & &  - & -  & &  - & -  & &  -0.0068 & 0.026
\\
\bottomrule
\end{tabular}
\end{ruledtabular}
\caption{
Means and standard deviations of the marginalized likelihoods. We observe that all additional DE parameters are consistent with $\Lambda$CDM at the $0.68\%$ C.L.\ for all models $w$CDM, $w$ZCDM1 as well as $w$ZCDM2. 
All models have unit rest frame sound speed, $\hat c_s=1$.
\label{tab_res}}
\end{table*}

We perform the analysis for the four following models: in $w$ZCDM2 we vary $\{w_0,$ $\epsilon_R$, $\epsilon_M\}$; in $w$ZCDM1 we vary $\{w_0$, $\epsilon\}$ and set $\epsilon_R=\epsilon_M=\epsilon$; in $w$CDM we fix $\epsilon=0$ in addition and, finally, in $\Lambda$CDM we fix $w_0=-1$ as well. On top of these DE parameters we vary the common set of six base parameters. These are the slope $n_s$ and amplitude $\log[10^{10}A_s]$ of the spectrum of primordial scalar curvature perturbations (modeled as a power law normalized at $k=0.05\,\text{Mpc}^{-1}$), the depth to re-ionization $\tau$, the physical baryon energy fraction $\Omega_bh^2$, the physical CDM energy fraction $\Omega_{\rm DM}h^2$, and 100 times the ratio of the  sound horizon to the angular diameter distance to the last-scattering surface, $\theta$. We use flat priors for all parameters, set adiabatic initial conditions for the evolution of the cosmological perturbations and ignore vector and tensor modes for simplicity.

\begin{figure*}[t]
\centering
\includegraphics[width=0.85\textwidth]{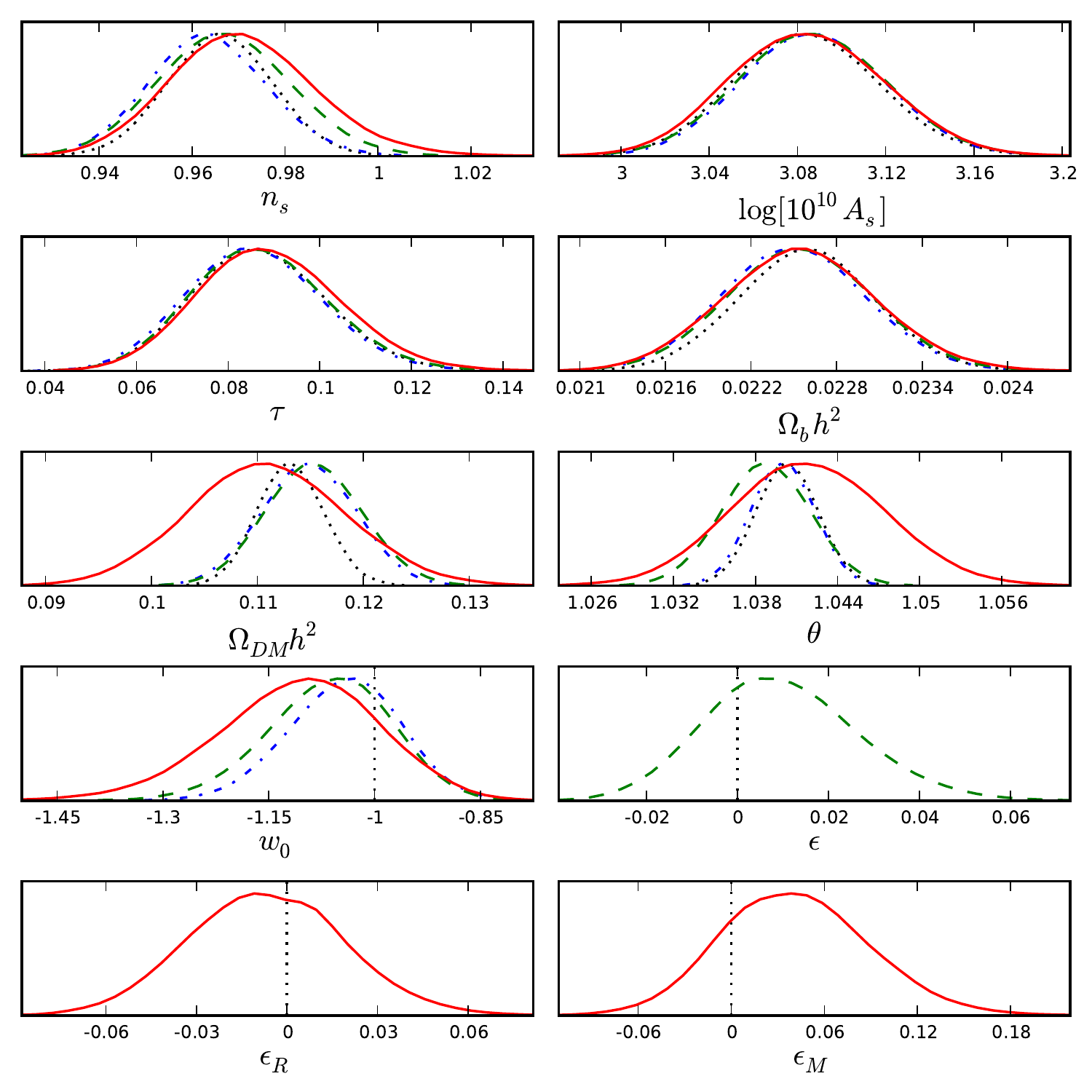}
\caption{\label{fig_1d_1}
Posterior probabilities after marginalization over the other parameters: $\Lambda$CDM (dotted black), $w$CDM (dot-dashed blue), $w$ZCDM1 (dashed green)  and $w$ZCDM2 (solid red); all models with unit rest frame sound speed, $\hat c_s=1$.}
\end{figure*}

As opposed to the cosmological constant $\Lambda$, in the case of dynamical DE models we need to model the evolution of its inhomogeneities. If we were interested in a specific model, we should solve the perturbation equations derived from its Lagrangian. However, as discussed in Section~\ref{sect:more}, in the context of this work we are not really interested in analyzing a model with a specific form of the functions $F(\chi)$ and $V(\chi)$. We rather wish to extract more general features of a class of DE models suggested by our theoretical considerations. In this spirit, we will at first assume the DE to be a perfect fluid with a constant rest frame sound speed $\hat c_s=1$, as it is usually done in the $w$CDM model, and we will later check how our results depend on this assumption.

The results are summarized in table \ref{tab_res}. We show the marginalized posterior probability distributions for $w$ZCDM1 and  $w$ZCDM2 in fig.~\ref{fig_1d_1}, comparing both cases with $w$CDM and $\Lambda$CDM. For $w$ZCDM1 we find lower and upper limits of 
\be
-1.25 < w_0 < -0.908\, ,\qquad 
-0.0201 < \epsilon < 0.0460\, ,
\ee
at 95\% C.L. For $w$ZCDM2 we find lower and upper limits of 
\bees
&&-1.35 < w_0 < -0.903\, , \qquad
-0.0465 < \epsilon_M < 0.140\, ,\nn\\
&&\hspace*{15mm}-0.0558 < \epsilon_R < 0.0446\, ,
\ees
again at 95\% C.L. In general, we observe a slight preference for models with $w_0<-1$ and $\eps_M>0$, i.e.\ early DE models with a recent phantom-crossing. However, all new parameters are consistent with their $\Lambda$CDM values at the $95\%$ C.L.\ as this is also the case for $w_0$ in $w$CDM. The results are consistent with those for other early DE models found in the literature, e.g.\ \cite{Xia:2009ys,dePutter:2010vy,Reichardt:2011fv}. However, note that our model has a slightly different evolution during radiation domination than the widely used parameterization proposed in ref.~\cite{Doran:2006kp}, and usually for that model phantom-crossing is excluded with the hard prior $w_0\geq -1$ which we do not employ here.
 
\begin{figure*}[t]
\centering
\includegraphics[width=0.85\textwidth]{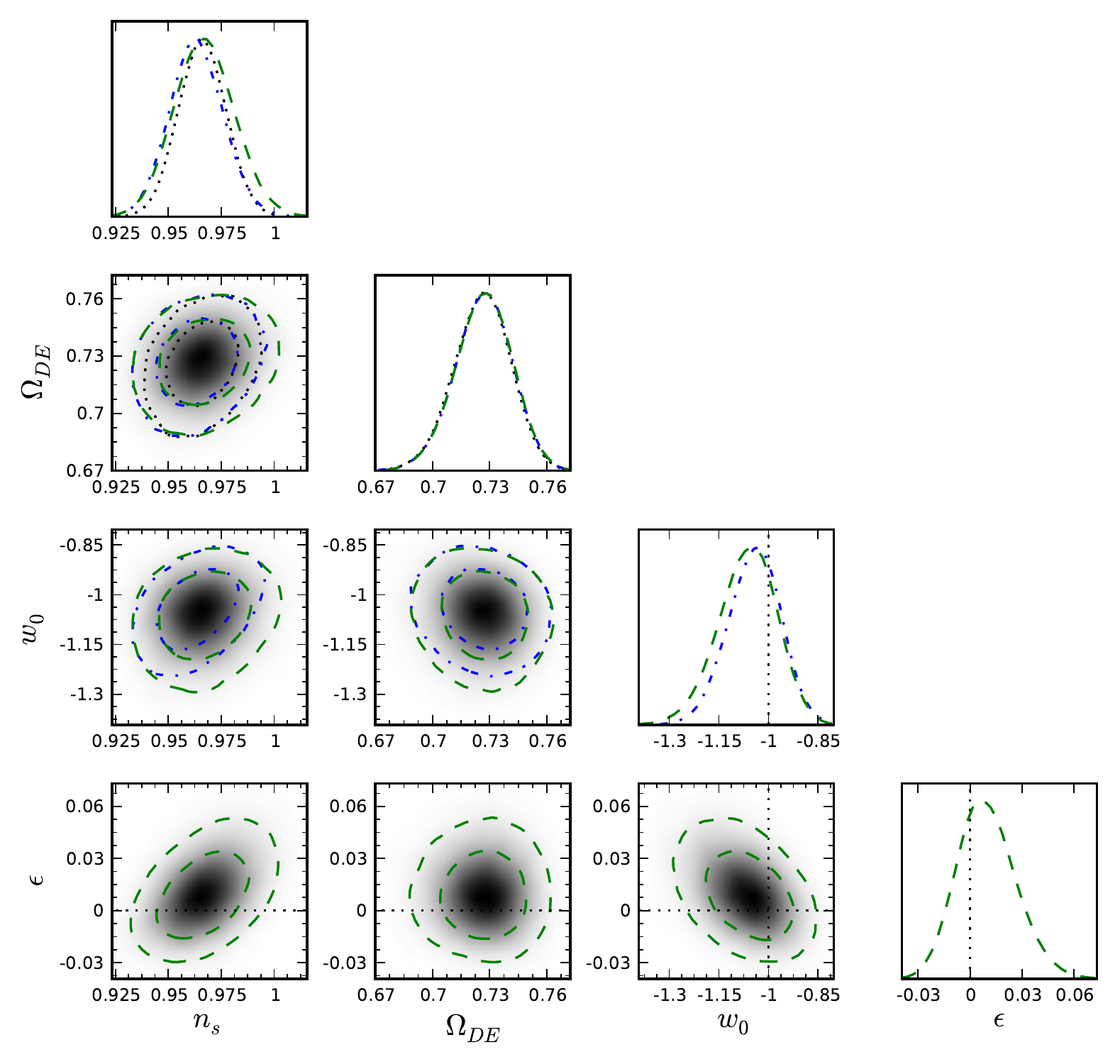} 
\caption{\label{fig_tri_1}
$w$ZCDM1: The marginalized likelihood contours at $1\sigma$ and $2\sigma$ are shown for all combinations of the parameters $n_s$, $\Omega_{\rm DE}$, $w_0$ and $\eps$ in $w$ZCDM1. We compare
$\Lambda$CDM (dotted black), $w$CDM (dot-dashed blue) and  $w$ZCDM1 (dashed green)
 The shading is the mean likelihood surface for the $w$ZCDM1 model. All models have unit rest frame sound speed, $\hat c_s=1$.}
\end{figure*}

In fig.~\ref{fig_tri_1} we show the combined likelihood contours for a subset of the parameters of $w$ZCDM1 and compare the constraints with those for the $w$CDM model. The slight degeneracy in the $(\eps,\,w_0)$ plane comes from the fact that $w_0(a)$ depends logarithmically on $\eps(a)$ as can be seen from \eq{defeps}. For positive $\eps$ the data prefers $w_0<-1$. Secondly, we note that the constraints on $\Omega_{\rm DE}$ are virtually unchanged by introducing the new parameter $\eps$, as there is basically no degeneracy in the $(\eps,\,\Omega_{\rm DE})$ plane and the marginalized posterior of for $\Omega_{\rm DE}$ is unchanged as well. Finally, $\eps$ is mildly correlated with the spectral index, $n_s$, as our analysis shows. This leads to a slight shift of the posterior to larger values of $n_s$ in $w$ZCDM1 as compared to $w$CDM.

\begin{figure*}[t]
\centering
\includegraphics[width=0.85\textwidth]{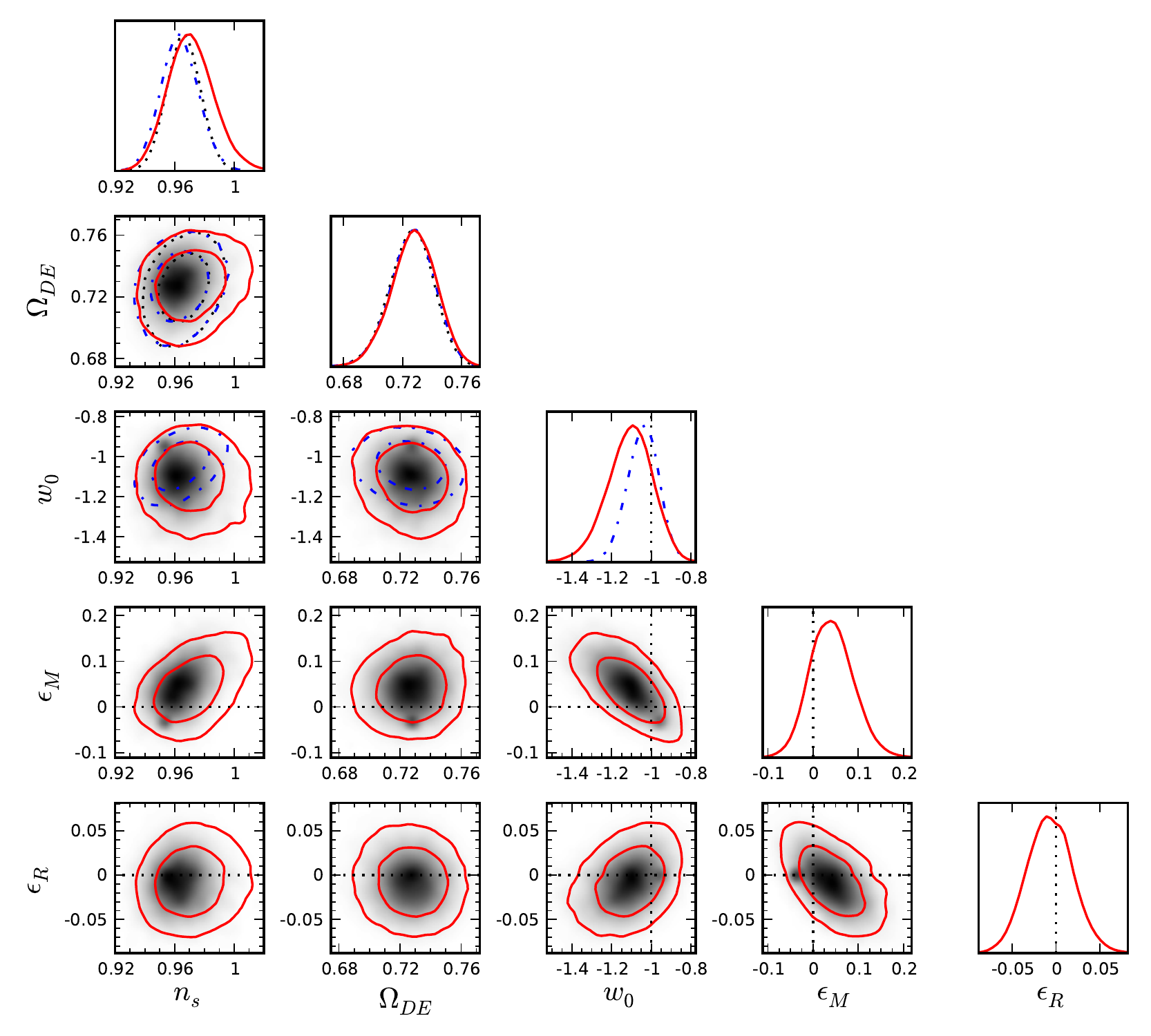} 
\caption{\label{fig_tri_2}
$w$ZCDM2: The marginalized likelihood contours at $1\sigma$ and $2\sigma$ are shown for all combinations of the parameters $n_s$, $\Omega_{\rm DE}$, $w_0$, $\eps_R$ and $\eps_M$. Red solid lines are for $w$ZCDM2 and blue dot-dashed lines are for $w$CDM. The shading is the mean likelihood surface for the $w$ZCDM2 model.  All models have unit rest frame sound speed, $\hat c_s=1$.}
\end{figure*}

In fig.~\ref{fig_tri_2} we show the combined likelihood contours for $w$ZCDM2,  comparing to $w$CDM. The correlation between $\eps_M$ and $w_0$ persists, which means that $\eps$ in the simplified $w$ZCDM1 model is more or less reflecting $\eps_M$. This is because the observations of the SNe and BAO are more sensitive to the late-time than the early Universe expansion history.  Introducing the additional parameter $\eps_R$ weakens the constraints on most parameters considerably. This is also a consequence of the CMB data not being strongly sensitive to $\eps(a)$. Therefore, other high-redshift probes such as 21~cm line spectra, gamma-ray bursts and Lyman-alpha forest data need to be added to improve the constraints for such models considerably.

\begin{figure*}[t]
\centering
\includegraphics[width=0.85\textwidth]{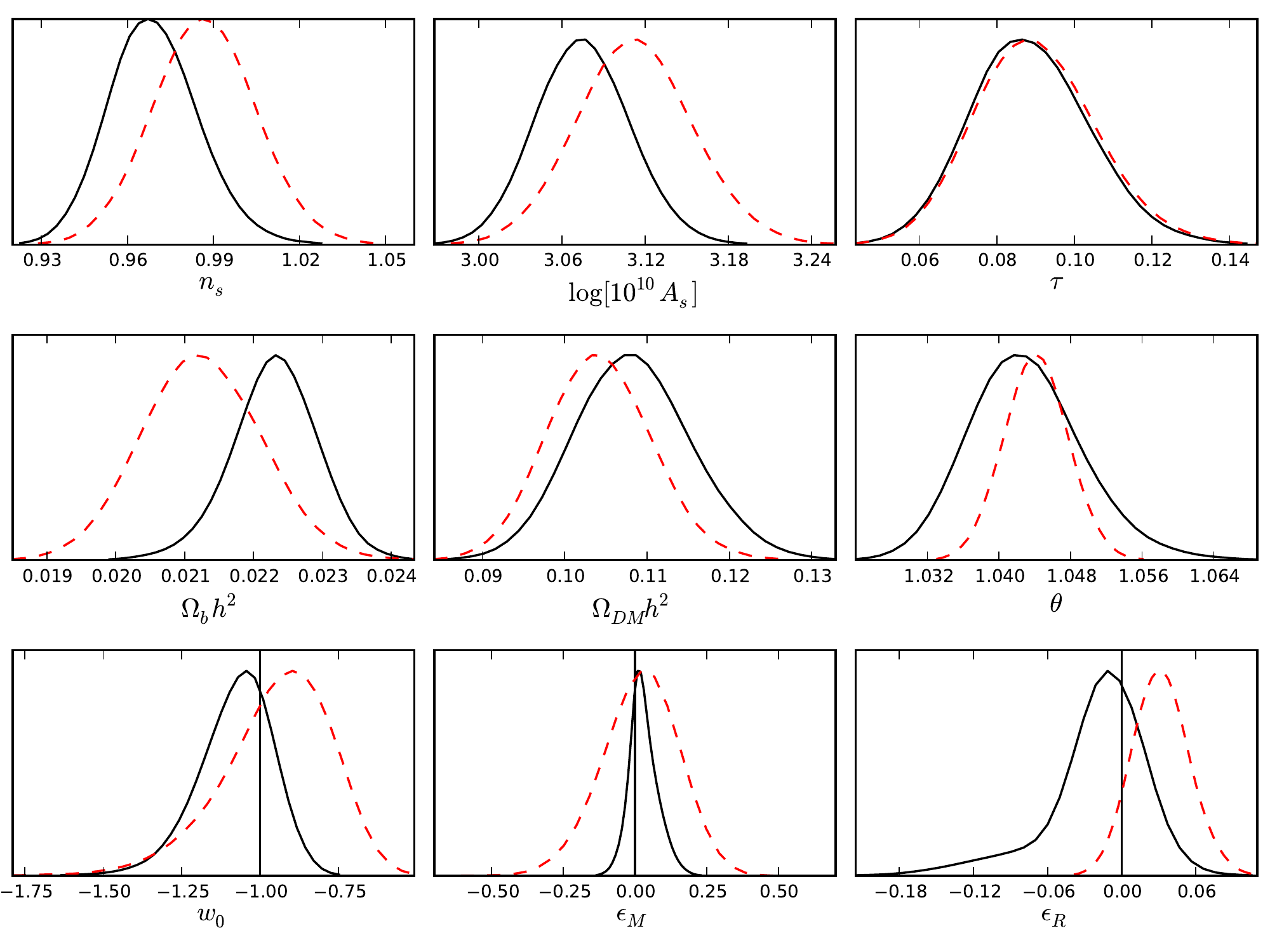} 
\caption{\label{ZCSe2wc1} 
Posterior probabilities after marginalization over the other parameters for $w$ZCDM2 with: $\hat{c}^2_s=1$ (solid black) and $\hat{c}^2_s=10^{-5}$ (dashed red). We note that for a small $\hat{c}_s$, the value of $\epsilon_M$ is less constrained, a positive value is slightly preferred for $\epsilon_R$ and that $w_0$ is less negative.}
\end{figure*}

We next examine the dependence of our results on the form of the perturbations.
It should be remarked that in Extended Quintessence (EQ) models, such as those represented by our 2-epsilons parametrization, the non-minimal coupling of the scalar field with the curvature leads to a specific behaviour of the perturbations, which is in general not equivalent to the one of a perfect fluid with unit rest-frame sound speed  $\hat{c}_s=1$. This issue was discussed in great details in refs.~\cite{Perrotta:2002sw} and \cite{Pettorino:2008ez}. In \cite{Perrotta:2002sw} it was found that, since the fluctuation in the energy density of the scalar field, $\delta \rho_{\chi}$, is directly related to the perturbation in the Ricci scalar, $\delta\rho_{\chi}$ tracks the density perturbation of the dominant component of the Universe. In MD, in particular, $\delta_ {\chi} \sim \delta_M$ on all sub-horizon scales, (where the density contrast is defined by $\delta_i=\delta \rho_i/ \delta \rho_i$) leading to the possibility of formation of dark energy clumps. In others words, in EQ $\hat{c}_s$ can drop much below unity, in contrast with what happens in normal quintessence where $\hat{c}_s=1$ such that dark energy perturbations are strongly suppressed on all sub-horizon scales. In principle, this can affect the growth of perturbations in the matter sector and give observable effects. In ref.~\cite{Perrotta:2002sw}, it was determined that values such as $\hat{c}_s^2\sim 0.001$ during MD are typical for EQ models. Moreover a small level of anisotropic stress is also present in EQ giving viscosity effects. 
In principle, the evolution of the perturbations in a given EQ model should be treated by solving the full system of perturbations equations for the specific choice of Lagrangian. 
Nevertheless, given the appropriate behaviour of the sound speed and the viscosity, the system can correctly be described in terms of fluid variables. 
To see how the constraints on our models depend on the sound speed we carry out the analysis for several values for $\hat{c}_s$, going as low as $10^{-5}$.
On the other hand, as shown in~\cite{Perrotta:2002sw}, the anisotropic stress in EQ models is typically small, and so we neglect it.
The results are shown in fig.~\ref{ZCSe2wc1}, where we compare the parameter estimation for $\hat{c}_s^2=1$ with that for 
the extreme case of $\hat{c}^2_s=10^{-5}$. This quantifies the dependence of the results on our assumptions on the evolution of the DE perturbations. We see that
for  small $\hat{c}_s$ the value of $\epsilon_M$ is less constrained, but in any case the bounds on $\eps_M$ and $\eps_R$ remain of the same order of magnitude. The fact that the mean of the posterior of $\eps_R$ shifts to the right when allowing DE to cluster is in line with the tendency of the CMB data to favour an additional clustering relativistic degree of freedom, see for instance \cite{Calabrese:2011hg}.
The results for $\hat{c}^2_s=10^{-3}$ are indistinguishable from those for $\hat{c}^2_s=10^{-5}$, so they are not shown. 
Thus, the model for the evolution of perturbations used in the previous analysis is sufficiently general, at least at the present level of accuracy of the data. This is in agreement with the results of
refs.~\cite{Sapone:2009mb,dePutter:2010vy,Ballesteros:2010ks,Sapone:2010uy} who indeed show that,
even for a small sound speed the effects are hardly detectable with current and even future data.

\section{Conclusions}

In this paper we have studied a number of issues related to vacuum fluctuations in a FRW background, and the role that they could  play  in explaining the observed acceleration of the Universe. 

First, we have studied vacuum fluctuations in a FRW background using a sharp cutoff in momentum space, and we have explicitly shown how one can recover the same results for the physical quantities that one obtains using a covariant regularization. The basic point is that, since a momentum-space cutoff breaks general covariance, 
one must allow for non-covariant counter-terms. A number of apparent pathologies discussed in the literature then disappear. In fact they are seen to be a consequence of 
a certain confusion between the bare and the renormalized $\cav\Tmn\vac$. While the bare energy density and pressure do not satisfy energy-momentum conservation, we can renormalize the theory so that  the renormalized quantities do satisfy it. We have also clarified some confusion about the EOS parameter of the term proportional to $H^2(t)$ in the vacuum energy density, for which general covariance requires $w=w_{\rm tot}(t)$ rather than $w=-1$.

Further, we have proposed an  alternative to the standard renormalization of vacuum energy, in which the $\lc^4$ divergence responsible for the cosmological constant problem is eliminated  at the classical level, following a procedure that is already part of the standard tenets of  classical general relativity and is used to define the  ADM mass of a space-time. The crucial point here is to realize that, even if it is true that in GR the choice of the zero of the energy is not arbitrary, the correct choice is still non-trivial, and requires  appropriate subtractions to be performed. We have seen that this point is further supported by the AdS/CFT correspondence, where the subtraction on the gravitational side corresponds to a renormalization  of UV divergences on the CFT side. This subtraction scheme provides a possible solution to the ``old" cosmological problem, namely why the vacuum energy density is not of order $\mpl^4$ nor of order $M^4$, where $M$ is larger than the TeV scale, up to which quantum field theory is successfully applied.

We then turned to the contribution to vacuum fluctuations  due to an ultra-light scalar field with mass $m<H_0$. We have seen that, in general, such a field generates a contribution of the form $f(\chi)R$ in the effective action (from which $\cav\Tmn\vac$ is derived by functional differentiation). Such a term leads to cosmological models where the DE density tracks the dominant energy density component, thereby  significantly alleviating the coincidence problem. We have performed a detailed comparison with cosmological observations of models where the DE density has the form suggested by the above theoretical considerations. We have seen that, quite generally, we can describe these models with a two-parameter extension of $w$CDM cosmology and, using a full likelihood analysis of CMB, BAO and SNIa data, we have presented the  limits on the parameters 
$\eps_R$ and $\eps_M$ that encode the deviations from $w$CDM. They are consistent with their $\Lambda$CDM values, as it is also the case for $w_0$ in $w$CDM. Thereby there is at present  no evidence for the presence of an ultralight scalar field that couples in this way to the curvature scalar.

\section*{Acknowledgements}
We thank Ramy Brustein, Claudia de~Rham, Rajeev Kumar Jain, Martin Kunz, 
Marcos Mari\~no, Alberto Nicolis, Valeria Pettorino, Massimo Porrati, Riccardo Rattazzi, Antonio Riotto and Misha Shaposhnikov for very useful discussions. Our work is supported by the Fonds National Suisse.

\appendix

\section{Non-covariant counter-term for the $H^2(t)\lc^2$ divergence}\label {app:noncov}

Considering again for illustration a massless minimally-coupled scalar field,
the bare energy density and pressure, as far as the term $\propto H^2(t)\lc^2$ are concerned,  can be written as 
\bees
\rho_{\rm bare}(\lc)&=&\frac{H^2(t)\lc^2}{16\pi^2}\, ,\label{rhobareH}\\
p_{\rm bare}(\lc)&=&\[w_{\rm tot}(t)+\frac{2}{3}\] \frac{H^2(t)\lc^2}{16\pi^2}
\, , \label{pbareH}
\ees
where we used \eq{c1w}. A counter-term proportional to the Einstein-Hilbert action,
\be
S_C=C(\lc)\int d^4x\sqrt{-g}\, R\, ,
\ee
produces a counter-term 
\be
\cav\Tmn \vac =-\frac{2}{\sqrt{-g}}\frac{\d S_c}{\d\gMN}=
-2C(\lc)\Gmn\, , 
\ee
where $\Gmn$ is the Einstein tensor. In FRW,  $G_{00}=3H^2$ and $G_{ij}=-a^2\d_{ij}(2\dot{H}+3H^2)=+3a^2\d_{ij} w_{\rm tot}(t)H^2(t)$, 
(where we used the relation
$2\dot{H}=-3 [1+w_{\rm tot}(t)] H^2$).
This gives a counter-term for the energy density 
\be
\rho_C=-6C(\lc)H^2(t)\, ,
\ee
and a counter-term for the pressure 
\be
p_C=-6C(\lc)w_{\rm tot}(t)H^2(t)\, . 
\ee
The function $C(\lc)$ can now be chosen so to cancel the divergence $H^2\lc^2/(16\pi^2)$ in \eq{rhobareH} simply choosing $6C(\lc)=\lc^2/(16\pi^2)+c_1$, with $c_1$ a finite constant which corresponds to a finite renormalization of $G$, and can therefore be simply set to zero. This
will also automatically cancel the divergence 
$w_{\rm tot}H^2(t)\lc^2/(16\pi^2)$ in \eq{pbareH}.

We then need a second counter-term which acts only on the pressure and cancels the term $(2/3)H^2(t)\lc^2/(16\pi^2)$ in \eq{pbareH}. To explicitly write the required non-covariant counter-term, it is convenient to work directly in a FRW space-time with scale factor $a(t)$. In order to perform the variation of the counter-term with respect to the FRW metric we introduce $\gmn =(-N^2(t), b^2(t)\d_{ij})$ and perform the variation with respect to $N(t)$ and $b(t)$. Then,  {\em after} performing the variation, we can set the lapse function $N(t)$ equal to one, and the function $b(t)$ equal to the desired background value $a(t)$. A counter-term $S_{\rm count}$ in the action induces a counter-term for the energy density given by
\be
\rho_{\rm count}=-\frac{2}{\sqrt{-g}}g_{00}\frac{\d S_{\rm count}}{\d g_{00}}
=-\frac{1}{a^3}\, \(\frac{\d S_{\rm count}}{\d N}\)_{N=1}\, ,
\ee
while for the pressure
\be\label{pcountFRW}
p_{\rm count}=\frac{1}{3a^2}\, 
\(\frac{\d S_{\rm count}}{\d b}\)_{b(t)=a(t)}\, .
\ee
The counter-term  that cancels the divergence  $(2/3)H^2(t)\lc^2/(16\pi^2)$ in \eq{pbareH} will necessarily be non-covariant, and can in principle depend on the function $a(t)$ both implicitly, due to the fact that the variation with respect to $b(t)$ is evaluated  in $b(t)=a(t)$, and explicitly. A counter-term that does the job is
\bees
S_{\rm count}&=&-\frac{\lc^2}{96\pi^2}\int d^4x \sqrt{-\tilde{g}}\,  (\tilde{R}+2\tilde{R}_{00})\, 
\frac{\sum_{i}g_{ii}}{\sum_{i}\tilde{g}_{ii}}\nn\\
&=&
-\frac{\lc^2}{16\pi^2}\, \int dt\,  \frac{\dot{a}^2}{a} b^2(t)\, .
\ees
where $\tilde{g}_{\mu\nu}=(-1,a^2\d_{ij})$, $\gmn=(-N^2,b^2\d_{ij})$ and $
\tilde{R}_{\mu\nu}$ and $\tilde{R}$ are the Ricci tensor and  Ricci scalar constructed from $\tilde{g}_{\mu\nu}$.
By construction $S_{\rm count}$ does not depend on $N(t)$, so it does not contribute to the energy density. Its contribution to the pressure, from \eq{pcountFRW}, is
$-\lc^2H^2(t)/(24\pi^2)$ and therefore cancels the term 
$(2/3) H^2(t)\lc^2/(16\pi^2)$ in \eq{pbareH}.


\bibliography{ZPQF_refs}

\begin{thebibliography}{71}
\expandafter\ifx\csname natexlab\endcsname\relax\def\natexlab#1{#1}\fi
\expandafter\ifx\csname bibnamefont\endcsname\relax
  \def\bibnamefont#1{#1}\fi
\expandafter\ifx\csname bibfnamefont\endcsname\relax
  \def\bibfnamefont#1{#1}\fi
\expandafter\ifx\csname citenamefont\endcsname\relax
  \def\citenamefont#1{#1}\fi
\expandafter\ifx\csname url\endcsname\relax
  \def\url#1{\texttt{#1}}\fi
\expandafter\ifx\csname urlprefix\endcsname\relax\def\urlprefix{URL }\fi
\providecommand{\bibinfo}[2]{#2}
\providecommand{\eprint}[2][]{\url{#2}}

\bibitem[{\citenamefont{Birrell and Davies}(1982)}]{Birrell:1982ix}
\bibinfo{author}{\bibfnamefont{N.}~\bibnamefont{Birrell}} \bibnamefont{and}
  \bibinfo{author}{\bibfnamefont{P.}~\bibnamefont{Davies}},
  \emph{\bibinfo{title}{{Quantum fields in curved space}}}
  (\bibinfo{publisher}{Cambridge University Press, 340 p},
  \bibinfo{year}{1982}).

\bibitem[{\citenamefont{Parker and Fulling}(1974)}]{Parker:1974qw}
\bibinfo{author}{\bibfnamefont{L.}~\bibnamefont{Parker}} \bibnamefont{and}
  \bibinfo{author}{\bibfnamefont{S.~A.} \bibnamefont{Fulling}},
  \bibinfo{journal}{Phys. Rev.} \textbf{\bibinfo{volume}{D9}},
  \bibinfo{pages}{341} (\bibinfo{year}{1974}).

\bibitem[{\citenamefont{Fulling et~al.}(1974)\citenamefont{Fulling, Parker, and
  Hu}}]{Fulling:1974pu}
\bibinfo{author}{\bibfnamefont{S.}~\bibnamefont{Fulling}},
  \bibinfo{author}{\bibfnamefont{L.}~\bibnamefont{Parker}}, \bibnamefont{and}
  \bibinfo{author}{\bibfnamefont{B.}~\bibnamefont{Hu}},
  \bibinfo{journal}{Phys.Rev.} \textbf{\bibinfo{volume}{D10}},
  \bibinfo{pages}{3905} (\bibinfo{year}{1974}).

\bibitem[{\citenamefont{Fulling and Parker}(1974)}]{Fulling:1974zr}
\bibinfo{author}{\bibfnamefont{S.~A.} \bibnamefont{Fulling}} \bibnamefont{and}
  \bibinfo{author}{\bibfnamefont{L.}~\bibnamefont{Parker}},
  \bibinfo{journal}{Annals Phys.} \textbf{\bibinfo{volume}{87}},
  \bibinfo{pages}{176} (\bibinfo{year}{1974}).

\bibitem[{\citenamefont{Peebles and Ratra}(2003)}]{Peebles:2002gy}
\bibinfo{author}{\bibfnamefont{P.~J.~E.} \bibnamefont{Peebles}}
  \bibnamefont{and} \bibinfo{author}{\bibfnamefont{B.}~\bibnamefont{Ratra}},
  \bibinfo{journal}{Rev. Mod. Phys.} \textbf{\bibinfo{volume}{75}},
  \bibinfo{pages}{559} (\bibinfo{year}{2003}), \eprint{astro-ph/0207347}.

\bibitem[{\citenamefont{Keski-Vakkuri and Sloth}(2003)}]{KeskiVakkuri:2003vj}
\bibinfo{author}{\bibfnamefont{E.}~\bibnamefont{Keski-Vakkuri}}
  \bibnamefont{and} \bibinfo{author}{\bibfnamefont{M.~S.} \bibnamefont{Sloth}},
  \bibinfo{journal}{JCAP} \textbf{\bibinfo{volume}{0308}}, \bibinfo{pages}{001}
  (\bibinfo{year}{2003}), \eprint{hep-th/0306070}.

\bibitem[{\citenamefont{Mangano}(2010)}]{Mangano:2010hw}
\bibinfo{author}{\bibfnamefont{G.}~\bibnamefont{Mangano}},
  \bibinfo{journal}{Phys.Rev.} \textbf{\bibinfo{volume}{D82}},
  \bibinfo{pages}{043519} (\bibinfo{year}{2010}), \eprint{1005.2758}.

\bibitem[{\citenamefont{Sloth}(2010)}]{Sloth:2010ti}
\bibinfo{author}{\bibfnamefont{M.~S.} \bibnamefont{Sloth}},
  \bibinfo{journal}{Int. J. Mod. Phys.} \textbf{\bibinfo{volume}{D19}},
  \bibinfo{pages}{2259} (\bibinfo{year}{2010}), \eprint{1005.3241}.

\bibitem[{\citenamefont{Padmanabhan}(2005)}]{Padmanabhan:2004qc}
\bibinfo{author}{\bibfnamefont{T.}~\bibnamefont{Padmanabhan}},
  \bibinfo{journal}{Class.Quant.Grav.} \textbf{\bibinfo{volume}{22}},
  \bibinfo{pages}{L107} (\bibinfo{year}{2005}), \eprint{hep-th/0406060}.

\bibitem[{\citenamefont{Maggiore}(2011)}]{Maggiore:2010wr}
\bibinfo{author}{\bibfnamefont{M.}~\bibnamefont{Maggiore}},
  \bibinfo{journal}{Phys. Rev.} \textbf{\bibinfo{volume}{D83}},
  \bibinfo{pages}{063514} (\bibinfo{year}{2011}), \eprint{1004.1782}.

\bibitem[{\citenamefont{Maggiore et~al.}(2011)\citenamefont{Maggiore,
  Hollenstein, Jaccard, and Mitsou}}]{Maggiore:2011hw}
\bibinfo{author}{\bibfnamefont{M.}~\bibnamefont{Maggiore}},
  \bibinfo{author}{\bibfnamefont{L.}~\bibnamefont{Hollenstein}},
  \bibinfo{author}{\bibfnamefont{M.}~\bibnamefont{Jaccard}}, \bibnamefont{and}
  \bibinfo{author}{\bibfnamefont{E.}~\bibnamefont{Mitsou}},
  \bibinfo{journal}{Phys.Lett.} \textbf{\bibinfo{volume}{B704}},
  \bibinfo{pages}{102} (\bibinfo{year}{2011}), \eprint{1104.3797}.

\bibitem[{\citenamefont{Barvinsky and Vilkovisky}(1985)}]{Barvinsky:1985an}
\bibinfo{author}{\bibfnamefont{A.~O.} \bibnamefont{Barvinsky}}
  \bibnamefont{and} \bibinfo{author}{\bibfnamefont{G.~A.}
  \bibnamefont{Vilkovisky}}, \bibinfo{journal}{Phys. Rept.}
  \textbf{\bibinfo{volume}{119}}, \bibinfo{pages}{1} (\bibinfo{year}{1985}).

\bibitem[{\citenamefont{Buchbinder et~al.}(1992)\citenamefont{Buchbinder,
  Odintsov, and Shapiro}}]{Buchbinder:1992rb}
\bibinfo{author}{\bibfnamefont{I.~L.} \bibnamefont{Buchbinder}},
  \bibinfo{author}{\bibfnamefont{S.~D.} \bibnamefont{Odintsov}},
  \bibnamefont{and} \bibinfo{author}{\bibfnamefont{I.~L.}
  \bibnamefont{Shapiro}}, \emph{\bibinfo{title}{{Effective action in quantum
  gravity}}} (\bibinfo{publisher}{Institute of Physics},
  \bibinfo{address}{Bristol, UK}, \bibinfo{year}{1992}), ISBN
  \bibinfo{isbn}{0-7503-0122-8}.

\bibitem[{\citenamefont{Mukhanov and Winitzki}(2007)}]{Mukhanov:2007zz}
\bibinfo{author}{\bibfnamefont{V.}~\bibnamefont{Mukhanov}} \bibnamefont{and}
  \bibinfo{author}{\bibfnamefont{S.}~\bibnamefont{Winitzki}},
  \emph{\bibinfo{title}{{Introduction to quantum effects in gravity}}}
  (\bibinfo{publisher}{Cambridge University Press},
  \bibinfo{address}{Cambridge}, \bibinfo{year}{2007}).

\bibitem[{\citenamefont{Shapiro}(2008)}]{Shapiro:2008sf}
\bibinfo{author}{\bibfnamefont{I.~L.} \bibnamefont{Shapiro}},
  \bibinfo{journal}{Class. Quant. Grav.} \textbf{\bibinfo{volume}{25}},
  \bibinfo{pages}{103001} (\bibinfo{year}{2008}), \eprint{0801.0216}.

\bibitem[{\citenamefont{Akhmedov}(2002)}]{Akhmedov:2002ts}
\bibinfo{author}{\bibfnamefont{E.~K.} \bibnamefont{Akhmedov}}
  (\bibinfo{year}{2002}), \eprint{hep-th/0204048}.

\bibitem[{\citenamefont{Bilic et~al.}(2012)\citenamefont{Bilic, Domazet, and
  Guberina}}]{Bilic:2011rj}
\bibinfo{author}{\bibfnamefont{N.}~\bibnamefont{Bilic}},
  \bibinfo{author}{\bibfnamefont{S.}~\bibnamefont{Domazet}}, \bibnamefont{and}
  \bibinfo{author}{\bibfnamefont{B.}~\bibnamefont{Guberina}},
  \bibinfo{journal}{Phys.Lett.} \textbf{\bibinfo{volume}{B707}},
  \bibinfo{pages}{221} (\bibinfo{year}{2012}), \eprint{1110.2393}.

\bibitem[{\citenamefont{Bilic}(2011)}]{Bilic:2011zm}
\bibinfo{author}{\bibfnamefont{N.}~\bibnamefont{Bilic}},
  \bibinfo{journal}{Phys.Rev.} \textbf{\bibinfo{volume}{D83}},
  \bibinfo{pages}{105003} (\bibinfo{year}{2011}), \eprint{1104.1349}.

\bibitem[{\citenamefont{Arnowitt et~al.}(1962)\citenamefont{Arnowitt, Deser,
  and Misner}}]{Arnowitt:1962hi}
\bibinfo{author}{\bibfnamefont{R.~L.} \bibnamefont{Arnowitt}},
  \bibinfo{author}{\bibfnamefont{S.}~\bibnamefont{Deser}}, \bibnamefont{and}
  \bibinfo{author}{\bibfnamefont{C.~W.} \bibnamefont{Misner}}, in
  \emph{\bibinfo{booktitle}{{Gravitation: an introduction to current
  research}}}, edited by
  \bibinfo{editor}{\bibfnamefont{L.}~\bibnamefont{Witten}}
  (\bibinfo{publisher}{{John Wiley \& Sons}}, \bibinfo{address}{New York,
  London}, \bibinfo{year}{1962}), chap.~\bibinfo{chapter}{7}, pp.
  \bibinfo{pages}{227--265}, \eprint{gr-qc/0405109}.

\bibitem[{\citenamefont{Poisson}(2004)}]{pois04}
\bibinfo{author}{\bibfnamefont{E.}~\bibnamefont{Poisson}},
  \emph{\bibinfo{title}{A Relativist's Toolkit. The Mathematics of Black-Hole
  Mechanics}} (\bibinfo{publisher}{Cambridge University Press},
  \bibinfo{year}{2004}).

\bibitem[{\citenamefont{Gibbons and Hawking}(1977)}]{Gibbons:1976ue}
\bibinfo{author}{\bibfnamefont{G.~W.} \bibnamefont{Gibbons}} \bibnamefont{and}
  \bibinfo{author}{\bibfnamefont{S.~W.} \bibnamefont{Hawking}},
  \bibinfo{journal}{Phys. Rev.} \textbf{\bibinfo{volume}{D15}},
  \bibinfo{pages}{2752} (\bibinfo{year}{1977}).

\bibitem[{\citenamefont{Brown and York}(1993)}]{Brown:1992br}
\bibinfo{author}{\bibfnamefont{J.~D.} \bibnamefont{Brown}} \bibnamefont{and}
  \bibinfo{author}{\bibfnamefont{J.~W.} \bibnamefont{York},
  \bibfnamefont{Jr.}}, \bibinfo{journal}{Phys. Rev.}
  \textbf{\bibinfo{volume}{D47}}, \bibinfo{pages}{1407} (\bibinfo{year}{1993}),
  \eprint{gr-qc/9209012}.

\bibitem[{\citenamefont{Brown et~al.}(1994)\citenamefont{Brown, Creighton, and
  Mann}}]{Brown:1994gs}
\bibinfo{author}{\bibfnamefont{J.~D.} \bibnamefont{Brown}},
  \bibinfo{author}{\bibfnamefont{J.}~\bibnamefont{Creighton}},
  \bibnamefont{and} \bibinfo{author}{\bibfnamefont{R.~B.} \bibnamefont{Mann}},
  \bibinfo{journal}{Phys. Rev.} \textbf{\bibinfo{volume}{D50}},
  \bibinfo{pages}{6394} (\bibinfo{year}{1994}), \eprint{gr-qc/9405007}.

\bibitem[{\citenamefont{Hawking and Horowitz}(1996)}]{Hawking:1995fd}
\bibinfo{author}{\bibfnamefont{S.~W.} \bibnamefont{Hawking}} \bibnamefont{and}
  \bibinfo{author}{\bibfnamefont{G.~T.} \bibnamefont{Horowitz}},
  \bibinfo{journal}{Class. Quant. Grav.} \textbf{\bibinfo{volume}{13}},
  \bibinfo{pages}{1487} (\bibinfo{year}{1996}), \eprint{gr-qc/9501014}.

\bibitem[{\citenamefont{Balasubramanian and
  Kraus}(1999)}]{Balasubramanian:1999re}
\bibinfo{author}{\bibfnamefont{V.}~\bibnamefont{Balasubramanian}}
  \bibnamefont{and} \bibinfo{author}{\bibfnamefont{P.}~\bibnamefont{Kraus}},
  \bibinfo{journal}{Commun. Math. Phys.} \textbf{\bibinfo{volume}{208}},
  \bibinfo{pages}{413} (\bibinfo{year}{1999}), \eprint{hep-th/9902121}.

\bibitem[{\citenamefont{Kraus et~al.}(1999)\citenamefont{Kraus, Larsen, and
  Siebelink}}]{Kraus:1999di}
\bibinfo{author}{\bibfnamefont{P.}~\bibnamefont{Kraus}},
  \bibinfo{author}{\bibfnamefont{F.}~\bibnamefont{Larsen}}, \bibnamefont{and}
  \bibinfo{author}{\bibfnamefont{R.}~\bibnamefont{Siebelink}},
  \bibinfo{journal}{Nucl. Phys.} \textbf{\bibinfo{volume}{B563}},
  \bibinfo{pages}{259} (\bibinfo{year}{1999}), \eprint{hep-th/9906127}.

\bibitem[{\citenamefont{Lau}(1999)}]{Lau:1999dp}
\bibinfo{author}{\bibfnamefont{S.~R.} \bibnamefont{Lau}},
  \bibinfo{journal}{Phys. Rev.} \textbf{\bibinfo{volume}{D60}},
  \bibinfo{pages}{104034} (\bibinfo{year}{1999}), \eprint{gr-qc/9903038}.

\bibitem[{\citenamefont{Mann}(1999)}]{Mann:1999pc}
\bibinfo{author}{\bibfnamefont{R.~B.} \bibnamefont{Mann}},
  \bibinfo{journal}{Phys. Rev.} \textbf{\bibinfo{volume}{D60}},
  \bibinfo{pages}{104047} (\bibinfo{year}{1999}), \eprint{hep-th/9903229}.

\bibitem[{\citenamefont{Emparan et~al.}(1999)\citenamefont{Emparan, Johnson,
  and Myers}}]{Emparan:1999pm}
\bibinfo{author}{\bibfnamefont{R.}~\bibnamefont{Emparan}},
  \bibinfo{author}{\bibfnamefont{C.~V.} \bibnamefont{Johnson}},
  \bibnamefont{and} \bibinfo{author}{\bibfnamefont{R.~C.} \bibnamefont{Myers}},
  \bibinfo{journal}{Phys. Rev.} \textbf{\bibinfo{volume}{D60}},
  \bibinfo{pages}{104001} (\bibinfo{year}{1999}), \eprint{hep-th/9903238}.

\bibitem[{\citenamefont{Brustein et~al.}(2002)\citenamefont{Brustein, Eichler,
  Foffa, and Oaknin}}]{Brustein:2000hh}
\bibinfo{author}{\bibfnamefont{R.}~\bibnamefont{Brustein}},
  \bibinfo{author}{\bibfnamefont{D.}~\bibnamefont{Eichler}},
  \bibinfo{author}{\bibfnamefont{S.}~\bibnamefont{Foffa}}, \bibnamefont{and}
  \bibinfo{author}{\bibfnamefont{D.~H.} \bibnamefont{Oaknin}},
  \bibinfo{journal}{Phys. Rev.} \textbf{\bibinfo{volume}{D65}},
  \bibinfo{pages}{105013} (\bibinfo{year}{2002}), \eprint{hep-th/0009063}.

\bibitem[{\citenamefont{Zhou et~al.}(2011)\citenamefont{Zhou, Yue, Yang, and
  Zou}}]{Zhou:2011um}
\bibinfo{author}{\bibfnamefont{K.}~\bibnamefont{Zhou}},
  \bibinfo{author}{\bibfnamefont{R.-H.} \bibnamefont{Yue}},
  \bibinfo{author}{\bibfnamefont{Z.-Y.} \bibnamefont{Yang}}, \bibnamefont{and}
  \bibinfo{author}{\bibfnamefont{D.-C.} \bibnamefont{Zou}}
  (\bibinfo{year}{2011}), \eprint{1110.0065}.

\bibitem[{\citenamefont{Frieman et~al.}(1995)\citenamefont{Frieman, Hill,
  Stebbins, and Waga}}]{Frieman:1995pm}
\bibinfo{author}{\bibfnamefont{J.~A.} \bibnamefont{Frieman}},
  \bibinfo{author}{\bibfnamefont{C.~T.} \bibnamefont{Hill}},
  \bibinfo{author}{\bibfnamefont{A.}~\bibnamefont{Stebbins}}, \bibnamefont{and}
  \bibinfo{author}{\bibfnamefont{I.}~\bibnamefont{Waga}},
  \bibinfo{journal}{Phys.Rev.Lett.} \textbf{\bibinfo{volume}{75}},
  \bibinfo{pages}{2077} (\bibinfo{year}{1995}), \eprint{astro-ph/9505060}.

\bibitem[{\citenamefont{Sahni and Habib}(1998)}]{Sahni:1998at}
\bibinfo{author}{\bibfnamefont{V.}~\bibnamefont{Sahni}} \bibnamefont{and}
  \bibinfo{author}{\bibfnamefont{S.}~\bibnamefont{Habib}},
  \bibinfo{journal}{Phys.Rev.Lett.} \textbf{\bibinfo{volume}{81}},
  \bibinfo{pages}{1766} (\bibinfo{year}{1998}), \eprint{hep-ph/9808204}.

\bibitem[{\citenamefont{Parker and Raval}(1999)}]{Parker:1999td}
\bibinfo{author}{\bibfnamefont{L.}~\bibnamefont{Parker}} \bibnamefont{and}
  \bibinfo{author}{\bibfnamefont{A.}~\bibnamefont{Raval}},
  \bibinfo{journal}{Phys.Rev.} \textbf{\bibinfo{volume}{D60}},
  \bibinfo{pages}{063512} (\bibinfo{year}{1999}), \eprint{gr-qc/9905031}.

\bibitem[{\citenamefont{Starobinsky}(1986)}]{Starobinsky:1986fx}
\bibinfo{author}{\bibfnamefont{A.~A.} \bibnamefont{Starobinsky}}, in
  \emph{\bibinfo{booktitle}{Field Theory, Quantum Gravity and Strings}}, edited
  by \bibinfo{editor}{\bibfnamefont{H.~J.} \bibnamefont{De~Vega}}
  \bibnamefont{and} \bibinfo{editor}{\bibfnamefont{N.}~\bibnamefont{Sanchez}}
  (\bibinfo{publisher}{Springer Verlag}, \bibinfo{year}{1986}), pp.
  \bibinfo{pages}{107--126}.

\bibitem[{\citenamefont{Morikawa}(1990)}]{Morikawa:1989xz}
\bibinfo{author}{\bibfnamefont{M.}~\bibnamefont{Morikawa}},
  \bibinfo{journal}{Phys.Rev.} \textbf{\bibinfo{volume}{D42}},
  \bibinfo{pages}{1027} (\bibinfo{year}{1990}).

\bibitem[{\citenamefont{Calzetta and Hu}(1995)}]{Calzetta:1995ys}
\bibinfo{author}{\bibfnamefont{E.}~\bibnamefont{Calzetta}} \bibnamefont{and}
  \bibinfo{author}{\bibfnamefont{B.}~\bibnamefont{Hu}},
  \bibinfo{journal}{Phys.Rev.} \textbf{\bibinfo{volume}{D52}},
  \bibinfo{pages}{6770} (\bibinfo{year}{1995}), \eprint{gr-qc/9505046}.

\bibitem[{\citenamefont{Uzan}(1999)}]{Uzan:1999ch}
\bibinfo{author}{\bibfnamefont{J.-P.} \bibnamefont{Uzan}},
  \bibinfo{journal}{Phys.Rev.} \textbf{\bibinfo{volume}{D59}},
  \bibinfo{pages}{123510} (\bibinfo{year}{1999}), \eprint{gr-qc/9903004}.

\bibitem[{\citenamefont{Amendola}(1999)}]{Amendola:1999qq}
\bibinfo{author}{\bibfnamefont{L.}~\bibnamefont{Amendola}},
  \bibinfo{journal}{Phys. Rev.} \textbf{\bibinfo{volume}{D60}},
  \bibinfo{pages}{043501} (\bibinfo{year}{1999}), \eprint{astro-ph/9904120}.

\bibitem[{\citenamefont{Chiba}(1999)}]{Chiba:1999wt}
\bibinfo{author}{\bibfnamefont{T.}~\bibnamefont{Chiba}},
  \bibinfo{journal}{Phys.Rev.} \textbf{\bibinfo{volume}{D60}},
  \bibinfo{pages}{083508} (\bibinfo{year}{1999}), \eprint{gr-qc/9903094}.

\bibitem[{\citenamefont{Perrotta et~al.}(2000)\citenamefont{Perrotta,
  Baccigalupi, and Matarrese}}]{Perrotta:1999am}
\bibinfo{author}{\bibfnamefont{F.}~\bibnamefont{Perrotta}},
  \bibinfo{author}{\bibfnamefont{C.}~\bibnamefont{Baccigalupi}},
  \bibnamefont{and}
  \bibinfo{author}{\bibfnamefont{S.}~\bibnamefont{Matarrese}},
  \bibinfo{journal}{Phys.Rev.} \textbf{\bibinfo{volume}{D61}},
  \bibinfo{pages}{023507} (\bibinfo{year}{2000}), \eprint{astro-ph/9906066}.

\bibitem[{\citenamefont{Chiba}(2001)}]{Chiba:2001xx}
\bibinfo{author}{\bibfnamefont{T.}~\bibnamefont{Chiba}},
  \bibinfo{journal}{Phys.Rev.} \textbf{\bibinfo{volume}{D64}},
  \bibinfo{pages}{103503} (\bibinfo{year}{2001}), \eprint{astro-ph/0106550}.

\bibitem[{\citenamefont{Chiba et~al.}(2010)\citenamefont{Chiba, Siino, and
  Yamaguchi}}]{Chiba:2010cy}
\bibinfo{author}{\bibfnamefont{T.}~\bibnamefont{Chiba}},
  \bibinfo{author}{\bibfnamefont{M.}~\bibnamefont{Siino}}, \bibnamefont{and}
  \bibinfo{author}{\bibfnamefont{M.}~\bibnamefont{Yamaguchi}},
  \bibinfo{journal}{Phys.Rev.} \textbf{\bibinfo{volume}{D81}},
  \bibinfo{pages}{083530} (\bibinfo{year}{2010}), \eprint{1002.2986}.

\bibitem[{\citenamefont{Wetterich}(1995)}]{Wetterich:1994bg}
\bibinfo{author}{\bibfnamefont{C.}~\bibnamefont{Wetterich}},
  \bibinfo{journal}{Astron. Astrophys.} \textbf{\bibinfo{volume}{301}},
  \bibinfo{pages}{321} (\bibinfo{year}{1995}), \eprint{hep-th/9408025}.

\bibitem[{\citenamefont{Caldera-Cabral
  et~al.}(2009)\citenamefont{Caldera-Cabral, Maartens, and
  Schaefer}}]{CalderaCabral:2009ja}
\bibinfo{author}{\bibfnamefont{G.}~\bibnamefont{Caldera-Cabral}},
  \bibinfo{author}{\bibfnamefont{R.}~\bibnamefont{Maartens}}, \bibnamefont{and}
  \bibinfo{author}{\bibfnamefont{B.~M.} \bibnamefont{Schaefer}},
  \bibinfo{journal}{JCAP} \textbf{\bibinfo{volume}{0907}}, \bibinfo{pages}{027}
  (\bibinfo{year}{2009}), \eprint{0905.0492}.

\bibitem[{\citenamefont{Valiviita et~al.}(2010)\citenamefont{Valiviita,
  Maartens, and Majerotto}}]{Valiviita:2009nu}
\bibinfo{author}{\bibfnamefont{J.}~\bibnamefont{Valiviita}},
  \bibinfo{author}{\bibfnamefont{R.}~\bibnamefont{Maartens}}, \bibnamefont{and}
  \bibinfo{author}{\bibfnamefont{E.}~\bibnamefont{Majerotto}},
  \bibinfo{journal}{Mon. Not. Roy. Astron. Soc.}
  \textbf{\bibinfo{volume}{402}}, \bibinfo{pages}{2355} (\bibinfo{year}{2010}),
  \eprint{0907.4987}.

\bibitem[{\citenamefont{Basilakos et~al.}(2009)\citenamefont{Basilakos,
  Plionis, and Sola}}]{Basilakos:2009wi}
\bibinfo{author}{\bibfnamefont{S.}~\bibnamefont{Basilakos}},
  \bibinfo{author}{\bibfnamefont{M.}~\bibnamefont{Plionis}}, \bibnamefont{and}
  \bibinfo{author}{\bibfnamefont{J.}~\bibnamefont{Sola}},
  \bibinfo{journal}{Phys. Rev.} \textbf{\bibinfo{volume}{D80}},
  \bibinfo{pages}{083511} (\bibinfo{year}{2009}), \eprint{0907.4555}.

\bibitem[{\citenamefont{Grande et~al.}(2011)\citenamefont{Grande, Sola,
  Basilakos, and Plionis}}]{Grande:2011xf}
\bibinfo{author}{\bibfnamefont{J.}~\bibnamefont{Grande}},
  \bibinfo{author}{\bibfnamefont{J.}~\bibnamefont{Sola}},
  \bibinfo{author}{\bibfnamefont{S.}~\bibnamefont{Basilakos}},
  \bibnamefont{and} \bibinfo{author}{\bibfnamefont{M.}~\bibnamefont{Plionis}},
  \bibinfo{journal}{JCAP} \textbf{\bibinfo{volume}{1108}}, \bibinfo{pages}{007}
  (\bibinfo{year}{2011}), \eprint{1103.4632}.

\bibitem[{\citenamefont{Basilakos et~al.}(2012)\citenamefont{Basilakos, Bauer,
  and Sola}}]{Basilakos:2011wm}
\bibinfo{author}{\bibfnamefont{S.}~\bibnamefont{Basilakos}},
  \bibinfo{author}{\bibfnamefont{F.}~\bibnamefont{Bauer}}, \bibnamefont{and}
  \bibinfo{author}{\bibfnamefont{J.}~\bibnamefont{Sola}},
  \bibinfo{journal}{JCAP} \textbf{\bibinfo{volume}{1201}}, \bibinfo{pages}{050}
  (\bibinfo{year}{2012}), \eprint{1109.4739}.

\bibitem[{\citenamefont{Will}(2005)}]{Will:2005va}
\bibinfo{author}{\bibfnamefont{C.~M.} \bibnamefont{Will}},
  \bibinfo{journal}{Living Rev.Rel.} \textbf{\bibinfo{volume}{9}},
  \bibinfo{pages}{3} (\bibinfo{year}{2005}), \eprint{gr-qc/0510072}.

\bibitem[{\citenamefont{Lewis and Bridle}(2002)}]{Lewis:2002ah}
\bibinfo{author}{\bibfnamefont{A.}~\bibnamefont{Lewis}} \bibnamefont{and}
  \bibinfo{author}{\bibfnamefont{S.}~\bibnamefont{Bridle}},
  \bibinfo{journal}{Phys. Rev.} \textbf{\bibinfo{volume}{D66}},
  \bibinfo{pages}{103511} (\bibinfo{year}{2002}), \eprint{astro-ph/0205436}.

\bibitem[{\citenamefont{Lewis and Bridle}()}]{Cosmomc}
\bibinfo{author}{\bibfnamefont{A.}~\bibnamefont{Lewis}} \bibnamefont{and}
  \bibinfo{author}{\bibfnamefont{S.}~\bibnamefont{Bridle}},
  \emph{\bibinfo{title}{{Cosmological MonteCarlo (CosmoMC)}}},
  \bibinfo{note}{publicly available Markov-Chain Monte-Carlo likelihood
  sampler: http://cosmologist.info/cosmomc/}.

\bibitem[{\citenamefont{Lewis et~al.}(2000)\citenamefont{Lewis, Challinor, and
  Lasenby}}]{Lewis:1999bs}
\bibinfo{author}{\bibfnamefont{A.}~\bibnamefont{Lewis}},
  \bibinfo{author}{\bibfnamefont{A.}~\bibnamefont{Challinor}},
  \bibnamefont{and} \bibinfo{author}{\bibfnamefont{A.}~\bibnamefont{Lasenby}},
  \bibinfo{journal}{Astrophys. J.} \textbf{\bibinfo{volume}{538}},
  \bibinfo{pages}{473} (\bibinfo{year}{2000}), \eprint{astro-ph/9911177}.

\bibitem[{\citenamefont{Lewis and Challinor}()}]{Camb}
\bibinfo{author}{\bibfnamefont{A.}~\bibnamefont{Lewis}} \bibnamefont{and}
  \bibinfo{author}{\bibfnamefont{A.}~\bibnamefont{Challinor}},
  \emph{\bibinfo{title}{{Code for Anisotropies in the Microwave Background
  (CAMB)}}}, \bibinfo{note}{publicly available CMB-Boltzmann code:
  http://www.camb.info/www.camb.info}.

\bibitem[{\citenamefont{Riess et~al.}(2009)}]{Riess:2009pu}
\bibinfo{author}{\bibfnamefont{A.~G.} \bibnamefont{Riess}}
  \bibnamefont{et~al.}, \bibinfo{journal}{Astrophys. J.}
  \textbf{\bibinfo{volume}{699}}, \bibinfo{pages}{539} (\bibinfo{year}{2009}),
  \eprint{0905.0695}.

\bibitem[{\citenamefont{Pisanti et~al.}(2008)}]{Pisanti:2007hk}
\bibinfo{author}{\bibfnamefont{O.}~\bibnamefont{Pisanti}} \bibnamefont{et~al.},
  \bibinfo{journal}{Comp. Phys. Commun.} \textbf{\bibinfo{volume}{178}},
  \bibinfo{pages}{956} (\bibinfo{year}{2008}), \eprint{0705.0290}.

\bibitem[{\citenamefont{Amanullah et~al.}(2010)}]{Amanullah:2010vv}
\bibinfo{author}{\bibfnamefont{R.}~\bibnamefont{Amanullah}}
  \bibnamefont{et~al.}, \bibinfo{journal}{Astrophys. J.}
  \textbf{\bibinfo{volume}{716}}, \bibinfo{pages}{712} (\bibinfo{year}{2010}),
  \eprint{1004.1711}.

\bibitem[{\citenamefont{Komatsu et~al.}(2011)}]{Komatsu:2010fb}
\bibinfo{author}{\bibfnamefont{E.}~\bibnamefont{Komatsu}} \bibnamefont{et~al.}
  (\bibinfo{collaboration}{WMAP}), \bibinfo{journal}{Astrophys. J. Suppl.}
  \textbf{\bibinfo{volume}{192}}, \bibinfo{pages}{18} (\bibinfo{year}{2011}),
  \eprint{1001.4538}.

\bibitem[{\citenamefont{Reichardt et~al.}(2009)}]{Reichardt:2008ay}
\bibinfo{author}{\bibfnamefont{C.~L.} \bibnamefont{Reichardt}}
  \bibnamefont{et~al.}, \bibinfo{journal}{Astrophys. J.}
  \textbf{\bibinfo{volume}{694}}, \bibinfo{pages}{1200} (\bibinfo{year}{2009}),
  \eprint{0801.1491}.

\bibitem[{\citenamefont{Sievers et~al.}(2007)}]{Sievers:2005gj}
\bibinfo{author}{\bibfnamefont{J.~L.} \bibnamefont{Sievers}}
  \bibnamefont{et~al.}, \bibinfo{journal}{Astrophys. J.}
  \textbf{\bibinfo{volume}{660}}, \bibinfo{pages}{976} (\bibinfo{year}{2007}),
  \eprint{astro-ph/0509203}.

\bibitem[{\citenamefont{Reid et~al.}(2010)}]{Percival:2009xn}
\bibinfo{author}{\bibfnamefont{B.~A.} \bibnamefont{Reid}} \bibnamefont{et~al.}
  (\bibinfo{collaboration}{SDSS}), \bibinfo{journal}{Mon. Not. Roy. Astron.
  Soc.} \textbf{\bibinfo{volume}{401}}, \bibinfo{pages}{2148}
  (\bibinfo{year}{2010}), \eprint{0907.1660}.

\bibitem[{\citenamefont{Xia and Viel}(2009)}]{Xia:2009ys}
\bibinfo{author}{\bibfnamefont{J.-Q.} \bibnamefont{Xia}} \bibnamefont{and}
  \bibinfo{author}{\bibfnamefont{M.}~\bibnamefont{Viel}},
  \bibinfo{journal}{JCAP} \textbf{\bibinfo{volume}{0904}}, \bibinfo{pages}{002}
  (\bibinfo{year}{2009}), \eprint{0901.0605}.

\bibitem[{\citenamefont{de~Putter et~al.}(2010)\citenamefont{de~Putter,
  Huterer, and Linder}}]{dePutter:2010vy}
\bibinfo{author}{\bibfnamefont{R.}~\bibnamefont{de~Putter}},
  \bibinfo{author}{\bibfnamefont{D.}~\bibnamefont{Huterer}}, \bibnamefont{and}
  \bibinfo{author}{\bibfnamefont{E.~V.} \bibnamefont{Linder}},
  \bibinfo{journal}{Phys. Rev.} \textbf{\bibinfo{volume}{D81}},
  \bibinfo{pages}{103513} (\bibinfo{year}{2010}), \eprint{1002.1311}.

\bibitem[{\citenamefont{Reichardt et~al.}(2011)\citenamefont{Reichardt,
  de~Putter, Zahn, and Hou}}]{Reichardt:2011fv}
\bibinfo{author}{\bibfnamefont{C.~L.} \bibnamefont{Reichardt}},
  \bibinfo{author}{\bibfnamefont{R.}~\bibnamefont{de~Putter}},
  \bibinfo{author}{\bibfnamefont{O.}~\bibnamefont{Zahn}}, \bibnamefont{and}
  \bibinfo{author}{\bibfnamefont{Z.}~\bibnamefont{Hou}} (\bibinfo{year}{2011}),
  \eprint{1110.5328}.

\bibitem[{\citenamefont{Doran and Robbers}(2006)}]{Doran:2006kp}
\bibinfo{author}{\bibfnamefont{M.}~\bibnamefont{Doran}} \bibnamefont{and}
  \bibinfo{author}{\bibfnamefont{G.}~\bibnamefont{Robbers}},
  \bibinfo{journal}{JCAP} \textbf{\bibinfo{volume}{0606}}, \bibinfo{pages}{026}
  (\bibinfo{year}{2006}), \eprint{astro-ph/0601544}.

\bibitem[{\citenamefont{Perrotta and Baccigalupi}(2002)}]{Perrotta:2002sw}
\bibinfo{author}{\bibfnamefont{F.}~\bibnamefont{Perrotta}} \bibnamefont{and}
  \bibinfo{author}{\bibfnamefont{C.}~\bibnamefont{Baccigalupi}},
  \bibinfo{journal}{Phys.Rev.} \textbf{\bibinfo{volume}{D65}},
  \bibinfo{pages}{123505} (\bibinfo{year}{2002}), \eprint{astro-ph/0201335}.

\bibitem[{\citenamefont{Pettorino and Baccigalupi}(2008)}]{Pettorino:2008ez}
\bibinfo{author}{\bibfnamefont{V.}~\bibnamefont{Pettorino}} \bibnamefont{and}
  \bibinfo{author}{\bibfnamefont{C.}~\bibnamefont{Baccigalupi}},
  \bibinfo{journal}{Phys.Rev.} \textbf{\bibinfo{volume}{D77}},
  \bibinfo{pages}{103003} (\bibinfo{year}{2008}), \eprint{0802.1086}.

\bibitem[{\citenamefont{Calabrese et~al.}(2011)\citenamefont{Calabrese,
  Huterer, Linder, Melchiorri, and Pagano}}]{Calabrese:2011hg}
\bibinfo{author}{\bibfnamefont{E.}~\bibnamefont{Calabrese}},
  \bibinfo{author}{\bibfnamefont{D.}~\bibnamefont{Huterer}},
  \bibinfo{author}{\bibfnamefont{E.~V.} \bibnamefont{Linder}},
  \bibinfo{author}{\bibfnamefont{A.}~\bibnamefont{Melchiorri}},
  \bibnamefont{and} \bibinfo{author}{\bibfnamefont{L.}~\bibnamefont{Pagano}},
  \bibinfo{journal}{Phys.Rev.} \textbf{\bibinfo{volume}{D83}},
  \bibinfo{pages}{123504} (\bibinfo{year}{2011}), \eprint{1103.4132}.

\bibitem[{\citenamefont{Sapone et~al.}(2009)\citenamefont{Sapone, Kunz, and
  Kunz}}]{Sapone:2009mb}
\bibinfo{author}{\bibfnamefont{D.}~\bibnamefont{Sapone}},
  \bibinfo{author}{\bibfnamefont{M.}~\bibnamefont{Kunz}}, \bibnamefont{and}
  \bibinfo{author}{\bibfnamefont{M.}~\bibnamefont{Kunz}},
  \bibinfo{journal}{Phys.Rev.} \textbf{\bibinfo{volume}{D80}},
  \bibinfo{pages}{083519} (\bibinfo{year}{2009}), \eprint{0909.0007}.

\bibitem[{\citenamefont{Ballesteros and
  Lesgourgues}(2010)}]{Ballesteros:2010ks}
\bibinfo{author}{\bibfnamefont{G.}~\bibnamefont{Ballesteros}} \bibnamefont{and}
  \bibinfo{author}{\bibfnamefont{J.}~\bibnamefont{Lesgourgues}},
  \bibinfo{journal}{JCAP} \textbf{\bibinfo{volume}{1010}}, \bibinfo{pages}{014}
  (\bibinfo{year}{2010}), \eprint{1004.5509}.

\bibitem[{\citenamefont{Sapone et~al.}(2010)\citenamefont{Sapone, Kunz, and
  Amendola}}]{Sapone:2010uy}
\bibinfo{author}{\bibfnamefont{D.}~\bibnamefont{Sapone}},
  \bibinfo{author}{\bibfnamefont{M.}~\bibnamefont{Kunz}}, \bibnamefont{and}
  \bibinfo{author}{\bibfnamefont{L.}~\bibnamefont{Amendola}},
  \bibinfo{journal}{Phys.Rev.} \textbf{\bibinfo{volume}{D82}},
  \bibinfo{pages}{103535} (\bibinfo{year}{2010}), \eprint{1007.2188}.

\end{thebibliography}

\end{document}